\documentclass[a4paper,fleqn,usenatbib]{mnras}

\usepackage{newtxtext,newtxmath}

\usepackage[T1]{fontenc}
\usepackage{ae,aecompl}

\usepackage{graphicx}	
\usepackage{amsmath}	
\usepackage{amssymb}	
\usepackage{epstopdf}
\usepackage{natbib}
\usepackage{multirow}
\usepackage{hhline}
\usepackage{verbatim}
\usepackage[normalem]{ulem}
\newcommand{\lir}{L_{\rm{IR}}}
\newcommand{\lco}{L'_{\rm{CO}}}

\newcommand{\aco}{\alpha_{\mathrm{CO}}}

\newcommand{\degree}{$^\mathrm{o}$}

\title[]{The Fine Line Between Normal and Starburst Galaxies}

\author[N. Lee et al.]{Nicholas Lee,$^{1}$ Kartik Sheth,$^{2,3}$ Kimberly S. Scott,$^{3,4}$ Sune Toft,$^{1}$  
\newauthor Georgios E.\ Magdis,$^{1, 5}$ Ivana Damjanov,$^{6}$ H. Jabran Zahid,$^{6}$ Caitlin M.\ Casey,$^{7}$ 
\newauthor Isabella Cortzen,$^{1}$ Carlos G\'omez Guijarro,$^{1}$ Alexander Karim,$^{8}$ Sarah K.\ Leslie,$^{9}$ 
\newauthor Eva Schinnerer$^{9}$
\\
\\
$^{1}$Dark Cosmology Centre, Niels Bohr Institute, University of Copenhagen, Juliane Maries Vej 30, 2100 Copenhagen, Denmark\\
$^{2}$National  Aeronautics and Space Administration Headquarters, 300 E Street SW, Washington DC 20024-3210, USA\\
$^{3}$North American ALMA Science Center, National Radio Astronomy Observatory, Charlottesville, VA 22903, USA\\
$^{4}$University of Virginia Health System, 1215 Lee Street, Charlottesville, VA 22903, USA\\
$^{5}$Institute for Astronomy, Astrophysics, Space Applications and Remote Sensing, National Observatory of Athens, GR-15236 Athens, Greece\\
$^{6}$Harvard-Smithsonian Center for Astrophysics, 60 Garden Street, Cambridge, MA 02138, USA\\
$^{7}$Department of Astronomy, The University of Texas at Austin, 2515 Speedway Boulevard Stop C1400, Austin, TX 78712, USA\\
$^{8}$Argelander-Institut f\"{u}r Astronomie, Universit\"{a}t Bonn, Auf dem H\"{u}gel 71, D-53121 Bonn, Germany\\
$^{9}$Max-Planck Institute for Astronomy, Koenigstuhl 17, D-69117 Heidelberg, Germany
}

\date{Accepted XXX. Received YYY; in original form ZZZ}

\pubyear{2016}

\begin{document}
\label{firstpage}
\pagerange{\pageref{firstpage}--\pageref{lastpage}}
\maketitle

\begin{abstract}
Recent literature suggests that there are two modes through which galaxies grow their stellar mass -  a normal mode characterized by quasi-steady star formation, and a highly efficient starburst mode possibly triggered by stochastic events such as galaxy mergers. While these differences are established for extreme cases, the population of galaxies between these two regimes is poorly studied and it is not clear where the transition between these two modes of star formation occurs. We utilize ALMA observations of the CO $J=3\rightarrow2$ line luminosity in a sample of 20 infrared luminous galaxies that lie in the intermediate range between normal and starburst galaxies at $z \sim 0.25$--0.65 in the COSMOS field to examine their gas content and star formation efficiency.  We compare these quantities to the galaxies' deviation from the well-studied ``main sequence'' correlation (MS) between star formation rate and stellar mass and find that at log$(SFR/SFR_{\mathrm{MS}}) \lesssim 0.6$, a galaxy's distance to the main sequence is primarily driven by increased gas content, and not a more efficient  star formation process.  
\end{abstract}




\section{Introduction}
A framework of star formation in galaxies has emerged where galaxies fall into three distinct classes - (i) quiescent galaxies, which are no longer forming new stars, (ii) ``normal'' star-forming galaxies that all form their stars in a similar process that leads to smooth star formation histories, and (iii) starburst galaxies that undergo a physically different star-formation process, possibly triggered by a stochastic event such as a galaxy merger. The lines between these three galaxy classes are blurry and not well-defined, especially between ``normal'' star forming galaxies and starbursts. 

A commonly used method to identify starburst galaxies arises from the ``main sequence'' of star-forming galaxies (MS). The MS describes the tight, empirical correlation between star formation rate ($SFR$) and stellar mass ($M_{*}$) that the majority of star-forming galaxies follow \citep{2007ApJ...660L..43N,2007ApJ...670..156D,2007A&A...468...33E,2007ApJS..173..267S}.  The MS is commonly described as a single power law of the form $SFR \propto M_{*}^{\beta}$, with $\beta = 0.7$--1.0 and the normalization evolving to higher values at increasing redshift \citep{2007ApJ...660L..43N,2014ApJS..214...15S}.  

The existence and tightness of this MS correlation is commonly interpreted as evidence that the vast majority of galaxies are building their stellar mass through similar quasi-steady processes, while outliers to this sequence are considered starbursts (SB) possibly undergoing a different mode of star formation  \citep[e.g.][]{2011A&A...533A.119E,2011ApJ...739L..40R,2012ApJ...747L..31S,2014ApJ...796...25S}. However, \citet{2014arXiv1406.5191K} find that stochastic star formation would also create a MS correlation, and that the existence of the MS does not place any constraints on the uniformity of star formation efficiencies. In order to actually probe the efficiency of star formation occurring in these galaxies, we must have measurements of the molecular gas mass, which provides the fuel for star formation. 

A galaxy's star formation efficiency (SFE $\equiv SFR/M_{\rm{gas}}$) measures how quickly the galaxy is converting its available supply of molecular gas into stars.  Galaxies with high star formation efficiency must also have correspondingly low gas depletion timescales ($\tau_{depl} \equiv M_{\rm{gas}} / SFR$), and so are only able to maintain their current rate of star formation for a short burst of time.  Thus, it would be expected that SB galaxies should have high SFE. 

Studies of the molecular gas content of galaxies find that star-forming galaxies follow a well-defined power-law correlation between star formation rate density and molecular gas surface density, known as the Kennicutt-Schmidt Law \citep[see review in][]{2012ARA&A..50..531K}.  However, recent studies suggest that there may instead be two relationships that define distinct modes of star formation, with galaxies on the MS following the KS Law and SB galaxies lying on a separate, parallel relationship with systematically higher SFE \citep{2007ApJ...671..303B,2010ApJ...714L.118D,2010MNRAS.407.2091G,2015ApJ...812L..23S}.  While starburst galaxies appear to be undergoing a separate mode of star formation, studies suggest that main sequence galaxies all have similar SFEs so that the width of the MS is mostly driven by variations in gas content \citep[e.g.][]{2012ApJ...760....6M,2010ApJ...714L.118D,2010MNRAS.407.2091G}. 

Molecular gas is often inferred from the luminosity in ISM cooling lines such as CO, which require extremely time intensive observations. As a result, most studies of molecular gas at high-$z$ have focused on either the most luminous and extreme starburst systems \citep[e.g.][]{2005MNRAS.359.1165G,2006ApJ...640..228T,2008ApJ...680..246T,2013MNRAS.429.3047B}, targeted surveys of galaxies on the main sequence \citep{2010ApJ...714L.118D,2010MNRAS.407.2091G}, or lensed galaxies \citep[e.g.][]{2004ApJ...614L...5S,2010A&A...518L..35I,2011ApJ...739L..32R}. This means that the observed differences between main sequence and SB galaxies may be exaggerated by the very different selection effects at work. This problem is exacerbated by the common practice of adopting two different values for the $\aco$ conversion factor between observed CO luminosity and molecular gas mass for SB and MS galaxy samples \citep[see review in][]{2013ARA&A..51..207B}. This bimodal conversion factor stems from studies that suggest the well-calibrated Milky Way $\aco$ value results in gas masses that are larger than the dynamical mass in local starburst systems \citep{1999AJ....117.2632B}, and so adopt a smaller value of $\aco$ for starburst systems. Unfortunately, the adoption of this bimodal conversion factor may introduce or exaggerate many of the differences observed between SB and MS galaxies, including the two different relationships in the $SFR-M_{\mathrm{gas}}$ plane, which may merge together when using a continuously varying, metallicity-dependent conversion factor \cite[e.g.][]{2011MNRAS.418..664N}. 

Thus, it is vital to study galaxies in the transition region between main sequence and starburst galaxies. By comparing these galaxies to existing main sequence and starburst galaxy samples, we can determine how wide the main sequence is, where the starburst mode begins, and what role gas content and star formation efficiency play in this transition. The answers to these questions will have important implications for possible star formation histories, halo accretion rates, and star formation timescales \citep[e.g.][]{2014arXiv1406.5191K,2014ApJ...796...25S,2016arXiv160607436M}

In this paper, we present a study of 20 Luminous Infrared Galaxies (LIRGs) in the COSMOS field at redshifts $z = 0.25$--0.65 that sit in the transition region between ``normal'' main-sequence galaxies and starburst galaxies. We avoid introducing artificial dichotomy in the data from a bimodal $\aco$ conversion factor and examine whether these galaxies on the upper envelope of the main sequence have elevated SFRs because of increased gas content (like MS galaxies) or enhanced SFE (like starbursts). In Section \ref{sec:obs} we present our new CO line observations from ALMA Cycle 1 and in Section \ref{sec:data} we present the complementary multi-wavelength data. In Section \ref{sec:lir_lco} we examine correlations between observed quantities and then we analyse morphologies in Section \ref{sec:morph} before studying how gas mass and star formation efficiency affect distance to the main sequence in Section \ref{sec:phys}. Finally, we discuss the implications of our results and offer some interpretation in Section \ref{sec:discussion}. 

When calculating rest-frame quantities, we use a cosmology with $\Omega_{m} = 0.28$, $\Lambda = 0.72$, and $H_{0} = 70$ km s$^{-1}$ Mpc$^{-1}$ \citep{2013ApJS..208...19H}.  A \citet{2003PASP..115..763C} Initial Mass Function (IMF) truncated at 0.1 and 100 $M_{\odot}$ is used when deriving SFRs and stellar masses.

\section{ALMA Observations and Imaging}
\label{sec:obs}

\subsection{Sample Selection}
\label{ssec:sample}

Our sample is drawn from a population of far-IR/submm-bright galaxies detected in deep {\it Herschel Space Observatory} PACS and SPIRE imaging of the 2\,deg$^2$ COSMOS field \citep{2010MNRAS.409...48R,2011A&A...532A..90L}. We select galaxies with 250\,\micron\ flux density $S_{\mathrm 250} > $ 9\,mJy at $z=$0.25-0.65, where the CO $J=3\rightarrow2$ line ($\nu_{\mathrm{rest}} = 345.796$\,GHz) is redshifted into ALMA Band 6. This selection allows us to identify LIRGs for which we can expect to detect CO with ALMA in Cycle 1 in a reasonable amount of time.

We further limit the sample to galaxies with SEDs that are best-fit with Sa-Sdm or starburst galaxy templates (to avoid elliptical/S0 galaxies and active galactic nuclei), and cut galaxies with (NUV - r$^+$) $>3.5$ to avoid quiescent galaxies \citep{2013A&A...556A..55I}, ensuring that we are only looking at star-forming galaxies. These criterion result in a parent sample of 2122 IR-bright, star-forming galaxies at $z=$0.25-0.65. To give us further confidence in matching the low resolution {\it Herschel} data to the optical to near-IR data, we select galaxies from this parent sample that are detected at 1.4\,GHz from the VLA observations of the COSMOS field \citep[341/2122,][]{2010ApJS..188..384S}, and choose from galaxies that have previously known, secure spectroscopic redshifts \citep[197/341, e.g.][]{2007ApJS..172...70L}. This radio preselection does not bias our sample in terms of stellar mass or SFR, so we can be assured that we are studying a representative sample of LIRGs.

\begin{figure*}
\begin{center}
\includegraphics[width=17.5cm]{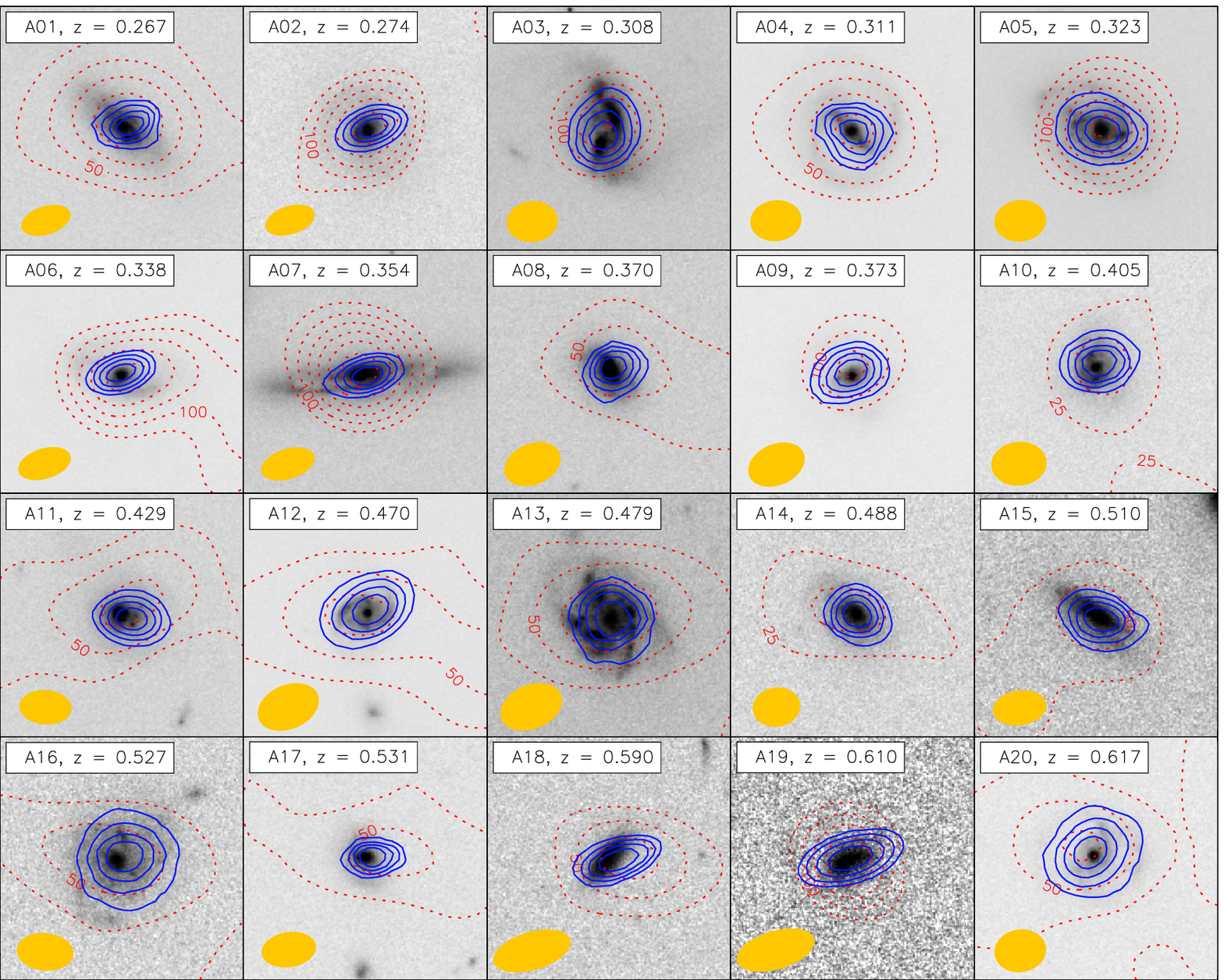}
\caption{{\it HST}/ACS F814W-band images for our sample of 20 $z =0.25$-0.65 LIRGs in the COSMOS field. The red contours show the VLA 1.4\,GHz continuum emission (in steps of 25\,$\mu$Jy), and the blue contours show the integrated line flux density (0.2, 0.4, 0.6, and 0.8 times the peak, in Jy\,km\,s$^{-1}$). Each image is 6\arcsec\ on a side. The yellow circle in the bottom-right of each image shows the synthesized beam for the ALMA observations.}
\label{fig:pstamps}
\end{center}
\end{figure*}

We finally select 20 galaxies for our ALMA observations to uniformly cover the desired redshift range -- with five galaxies in four redshift bins from $0.26 \le z < 0.33$, $0.33 \le z < 0.41$, $0.41 \le z < 0.52$, and $0.52 \le z < 0.62$ -- while optimizing and simplifying the correlator setup (see Section~\ref{ssec:almaobs}). The properties of these 20 targets are given in Table~\ref{tab:targets}, and ACS F814-W-band images from the {\it Hubble Space Telescope} ({\it HST}) are shown in Figure~\ref{fig:pstamps}.

\begin{table*}
\centering
\caption{LIRGs at $z = 0.25$-0.65 from the COSMOS field}
\label{tab:targets}
\begin{tabular}{ccccccccc} 
\hline
\hline
ALMA ID & Optical ID$^{\mathrm{a}}$ & $z_{\mathrm{spec}}$$^{\mathrm{b}}$ & log$(M_\ast)$$^{\mathrm{b}}$ & $S_{\mathrm{250}}$$^{\mathrm{b}}$ &
log$(L_{\mathrm{IR}})$$^{\mathrm{b}}$ & log$(M_{\mathrm{dust}})$$^{\mathrm{b}}$ & SFR$^{\mathrm{b}}$ & $\Delta SFR_{MS}$$^{\mathrm{b}}$ \\
 &  &  & log($M_\odot$) & mJy & log($L_\odot$) & log($M_\odot$) & $M_\odot$\,yr$^{-1}$ & log($\frac{SFR}{SFR_{\mathrm{MS}}}$) \\
\hline
ALMA01 & COSOPT J095936.95+014659.8 & 0.267 & $10.4 \pm 0.03$ & $30.4 \pm 1.07$ & $11.1 \pm 0.03$ & $ 8.1 \pm 0.09$ & $ 9.9 \pm 0.59$ & $ 0.2 \pm 0.03$\\ 
ALMA02 & COSOPT J100245.44+024049.7 & 0.274 & $10.3 \pm 0.03$ & $28.6 \pm 8.57$ & $11.4 \pm 0.02$ & $ 8.2 \pm 0.09$ & $19.6 \pm 0.77$ & $ 0.6 \pm 0.02$\\ 
ALMA03 & COSOPT J100139.09+021644.7 & 0.308 & $10.4 \pm 0.03$ & $35.6 \pm 1.10$ & $11.3 \pm 0.01$ & $ 8.1 \pm 0.05$ & $19.9 \pm 0.53$ & $ 0.5 \pm 0.02$\\ 
ALMA04 & COSOPT J100005.32+020135.4 & 0.311 & $10.6 \pm 0.04$ & $12.6 \pm 5.50$ & $11.1 \pm 0.03$ & $ 8.2 \pm 0.20$ & $10.1 \pm 0.76$ & $ 0.1 \pm 0.03$\\ 
ALMA05 & COSOPT J100036.34+015609.4 & 0.323 & $10.9 \pm 0.03$ & $45.2 \pm 10.96$ & $11.4 \pm 0.03$ & $ 8.4 \pm 0.20$ & $24.2 \pm 1.43$ & $ 0.5 \pm 0.03$\\ 
ALMA06 & COSOPT J095821.78+024820.8 & 0.338 & $11.1 \pm 0.02^{\mathrm{c}}$ & $29.0 \pm 3.02$ & $11.4 \pm 0.03$ & $ 8.1 \pm 0.16$ & $20.7 \pm 1.50$ & $ 0.3 \pm 0.03$\\ 
ALMA07 & COSOPT J095911.90+020824.2 & 0.354 & $11.1 \pm 0.02^{\mathrm{c}}$ & $38.0 \pm 12.44$ & $11.4 \pm 0.02$ & $ 8.5 \pm 0.14$ & $24.3 \pm 1.16$ & $ 0.4 \pm 0.02$\\ 
ALMA08 & COSOPT J100223.36+022642.6 & 0.370 & $10.1 \pm 0.06$ & $11.2 \pm 1.01$ & $11.1 \pm 0.12$ & $ 7.6 \pm 0.13$ & $12.7 \pm 3.41$ & $ 0.4 \pm 0.12$\\ 
ALMA09 & COSOPT J100138.75+022126.7 & 0.373 & $10.5 \pm 0.04$ & $22.7 \pm 1.43$ & $11.2 \pm 0.03$ & $ 8.1 \pm 0.14$ & $15.9 \pm 0.92$ & $ 0.3 \pm 0.03$\\ 
ALMA10 & COSOPT J100035.61+014502.5 & 0.405 & $10.5 \pm 0.03$ & $20.0 \pm 1.12$ & $11.3 \pm 0.20$ & $ 8.2 \pm 0.23$ & $16.7 \pm 7.41$ & $ 0.3 \pm 0.19$\\ 
ALMA11 & COSOPT J095906.68+014410.5 & 0.429 & $10.6 \pm 0.03$ & $19.4 \pm 1.11$ & $11.3 \pm 0.06$ & $ 8.1 \pm 0.17$ & $16.3 \pm 2.32$ & $ 0.2 \pm 0.06$\\ 
ALMA12 & COSOPT J095925.69+023609.7 & 0.470 & $10.4 \pm 0.04$ & $25.9 \pm 1.05$ & $11.5 \pm 0.03$ & $ 8.3 \pm 0.09$ & $27.5 \pm 1.55$ & $ 0.5 \pm 0.03$\\ 
ALMA13 & COSOPT J095904.44+014812.2 & 0.479 & $10.6 \pm 0.04$ & $28.1 \pm 1.03$ & $11.5 \pm 0.02$ & $ 8.3 \pm 0.07$ & $32.6 \pm 1.38$ & $ 0.5 \pm 0.02$\\ 
ALMA14 & COSOPT J095910.49+020618.0 & 0.488 & $10.5 \pm 0.05$ & $19.2 \pm 1.87$ & $11.4 \pm 0.04$ & $ 8.2 \pm 0.14$ & $20.3 \pm 1.74$ & $ 0.3 \pm 0.04$\\ 
ALMA15 & COSOPT J100138.14+020205.0 & 0.510 & $10.7 \pm 0.03$ & $19.6 \pm 1.05$ & $11.4 \pm 0.03$ & $ 8.2 \pm 0.11$ & $23.3 \pm 1.42$ & $ 0.3 \pm 0.03$\\ 
ALMA16 & COSOPT J100100.78+014430.8 & 0.527 & $10.6 \pm 0.04$ & $14.9 \pm 1.31$ & $11.3 \pm 0.12$ & $ 8.6 \pm 0.14$ & $18.0 \pm 4.34$ & $ 0.2 \pm 0.11$\\ 
ALMA17 & COSOPT J100134.34+020001.5 & 0.531 & $10.6 \pm 0.03$ & $13.7 \pm 1.87$ & $11.4 \pm 0.03$ & $ 7.8 \pm 0.10$ & $23.0 \pm 1.44$ & $ 0.3 \pm 0.03$\\ 
ALMA18 & COSOPT J100210.88+015429.6 & 0.590 & $10.8 \pm 0.04$ & $18.0 \pm 0.96$ & $11.5 \pm 0.02$ & $ 8.1 \pm 0.05$ & $28.0 \pm 1.00$ & $ 0.2 \pm 0.02$\\ 
ALMA19 & COSOPT J100128.11+024939.5 & 0.610 & $10.0 \pm 0.04$ & $26.6 \pm 1.03$ & $11.6 \pm 0.02$ & $ 8.5 \pm 0.07$ & $37.2 \pm 1.65$ & $ 0.8 \pm 0.04$\\ 
ALMA20 & COSOPT J095853.15+023617.9 & 0.617 & $10.7 \pm 0.04$ & $18.0 \pm 1.05$ & $11.3 \pm 0.08$ & $ 8.6 \pm 0.22$ & $22.1 \pm 3.26$ & $ 0.2 \pm 0.07$\\ 
\hline
\end{tabular}
\begin{flushleft}
$^{\mathrm{a}}$COSMOS identifications from \citet{2013A&A...556A..55I}.\\
$^{\mathrm{b}}$See Sections \ref{sec:obs} \& \ref{sec:cosmos} for derivations of galaxy properties.\\
$^{\mathrm{c}}$The listed stellar mass is the sum of the stellar masses of all galaxies within the {\em Herschel} SPIRE 250 $\mu$m beam (see Section \ref{sec:morph})\\
\end{flushleft}
\end{table*}

\subsection{ALMA Observations}
\label{ssec:almaobs}

These Cycle 1 data were carried out between January 1 to February 25 2014. Several calibration sources were used depending on availability at the time of observations. The bright quasars J1058+0133, J1037-2934, J0538-4405, and J0909+0121 are flat spectrum sources and were used as bandpass calibrators. The quasars J1008+0621 and J1058+0133 are close in proximity to our science targets ($\sim$5\degree, and $\sim$15\degree, respectively) and were used as phase calibrators. The original flux calibrators selected for our Cycle 1 observations were Solar system bodies including moons (Titan and Ganymede) and asteroids (Ceres and Pallas), but these objects have since been found to suffer from inaccurate absolute flux-density calibrations by the Nordic Regional ALMA Centre (Nordic ARC). Thus, following instruction by the Nordic ARC, the flux density scale for our observations was ultimately determined from the phase calibrators (see Section 2.3). 

The CO $J=3\rightarrow2$ transition falls in ALMA Band 6 for our sample of $z = 0.25$-0.65 LIRGs. Our targets were observed in single-pointing mode using the most compact configuration available in Cycle 1, resulting in low spatial resolution appropriate for detection experiments ($\theta_{\mathrm{FWHM}} \sim 1.3$\arcsec). We required only coarse spectral resolution, and used the TDM correlator mode (2000\,MHz bandwidth, 31.25\,MHz effective resolution), which gives a velocity resolution of 34-44\,km\,s$^{-1}$.

Given the wide range in redshifts for our targets, the most efficient observing strategy (i.e. that minimises calibration overheads) is to manually select the sky frequencies for the four spectral windows within each of the ALMA Science Goals. We arranged our 20 targets into five separate Science Goals such that multiple sources can be observed using the same spectral setup. Spectral windows that were not used for line detections were set to increase the total aggregate bandwidth available for the potential detection of the dust continuum.

Table~\ref{tab:almaobs} provides a summary of the ALMA observations, including the total on-source time and line sensitivities (rms) achieved (see Section~\ref{ssec:almaimage}). For each of the five Science Goals we used the galaxy with the lowest expected line flux density to set our sensitivity requirements. Due to excellent weather conditions, our achieved rms was often better than expected.

\begin{table*}
\centering
\caption{Summary of ALMA Observations}
\label{tab:almaobs}
\begin{tabular}{ccccccc} 
\hline
\hline
Science Goal & Targets & Dates Observed & PWV$^{\mathrm{a}}$ & On-source Time$^\mathrm{b}$ & $\theta_{\mathrm{maj}}$$^\mathrm{c}$ & $\theta_{\mathrm{min}}$$^\mathrm{c} $ \\
 &  &  & (mm) & (min) & (\arcsec) & (\arcsec) \\
\hline
CO(3-2)SG1 & A01,A02,A06,A07 & 25-Feb & 0.8 & 6.6 & 1.3 & 0.7 \\ 
CO(3-2)SG2 & A03,A04,A05,A10 & 11-Jan & 2.7 & 9.2 & 1.3 & 1.0  \\ 
CO(3-2)SG3 & A08,A09,A12,A13 & 11-Jan & 1.8 & 8.1 & 1.5 & 1.0 \\ 
CO(3-2)SG4a$^\mathrm{d}$ & A11,A16 & 14-Jan, 25-Feb & 3.9, 0.8 & 34.2 & 1.3 & 0.9 \\ 
CO(3-2)SG4b$^\mathrm{d}$ & A15,A17 & 25-Feb & 0.8 & 19.7 & 1.3 & 0.9 \\ 
CO(3-2)SG5a$^\mathrm{d}$ & A14,A20 & 01-Jan & 2.6, 3.1 & 48.3 & 1.2 & 1.0  \\ 
CO(3-2)SG5b$^\mathrm{d}$ & A18,A19 & 28-Jan, 25-Feb & 0.7, 0.8 & 45.2 & 2.0 & 0.9 \\ 
\hline
\end{tabular}
\begin{flushleft}
$^{\mathrm{a}}$Median precipitable water vapor (PWV) during observation. Multiple values indicate the median PWV during multiple executions of the Science Goal.\\
$^{\mathrm{b}}$On-source integration time per target.\\
$^{\mathrm{c}}$Synthesized beam width (FWHM) for the major ($\theta_{\mathrm{maj}}$) and minor ($\theta_{\mathrm{min}}$) axes.\\
$^{\mathrm{d}}$Both of these Science Goals were split into two during observations in order to restrict the total execution time as per standard ALMA operations.\\
\end{flushleft}
\end{table*}

\subsection{ALMA Imaging}
\label{ssec:almaimage}

These data were calibrated and imaged using CASA Version 4.7. Calibration of these data was carried out by ALMA staff at the Nordic Regional ALMA Centre using standard procedures to (1) determine and apply water vapor radiometer (WVR) and $T_{\mathrm{sys}}$ corrections, (2) identify and flag bad data, and (3) calculate and apply bandpass and phase solutions.  The absolute flux-density scale of the visibilities was computed by the
CASA task ``fluxscale''.  This task takes both the amplitude gain and the absolute flux density of at least one calibrator, scales the gains of these flux-density calibrators, and then uses these scales to derive the flux densities of the remaining calibrators. These gains are then interpolated to the target visibilities to apply an appropriate absolute flux-density scale.

In ordinary ALMA observations, the calibrator(s) with known flux
densities can be either solar-system objects or quasars monitored by
ALMA. In the former case, CASA estimates the flux density from the
sub-solar illumination, using an ephemeris table. In the latter case, we
can use the special ``getALMAflux'' function from the ARC analysis
utils\footnote{\url{https://casaguides.nrao.edu/index.php?title=Analysis_Utilities}}, to estimate the flux density of the quasar from an interpolation of the
multi-frequency ALMA monitoring.

Since several solar-system objects (in particular, asteroids such as 
Pallas) have shown to not give an
accurate flux-density calibration (Nordic ALMA Regional Centre Node, priv. comm.) and our
phase and bandpass calibrators happen to be included in the ALMA quasar
monitoring, we opted to perform our absolute flux-density calibration
using the ALMA quasar monitoring. This way, we also ensure a more
homogeneous and self-consistent calibration among the different
executions of our observations.

We imaged each science target using the CLEAN task in CASA. We imaged the spectral line data using Cotton-Schwab cleaning in frequency mode, using the native spectral resolution of 31.25\,MHz. We used the Hogbom method for PSF calculation, and natural weighting in order to maximize the signal-to-noise (S/N). We set the pixel size for each image such that there are five pixels per beam (minor axis; 0.13-0.20\arcsec), and set the image size to cover out to where the beam response falls to 0.5 (maps are either 256$\times$256 or 216$\times$216 pixels$^2$).

We use an iterative method to clean the data, as follows: (1) make the dirty image and identify which channels contain the line emission; (2) calculate the map rms ($\sigma$) from the channels with line emission; (3) run CLEAN again in interactive mode, drawing clean boxes to select only the line emission and cleaning down to a threshold of $1\sigma$; and (4) repeat steps 2 and 3, recalculating $\sigma$ each time, until it decreases by $\le$5\%. CO is detected in all 20 of our targets with high significance (Section~\ref{ssec:covalues}; see Figure~\ref{fig:pstamps}).

We also made continuum images using multi-frequency synthesis mode in CLEAN and all four spectral windows, excluding channels containing the CO $J=3\rightarrow2$ line. We used a similar iterative method to compute the rms and clean the continuum data, cleaning down to $1\sigma$ on continuum sources with flux density $\ge4\sigma$.  Only 7/20 of our targets are detected above $3\sigma$ at 1\,mm. 

\subsection{CO Intensities}
\label{ssec:covalues}

All 20 of the LIRGs are detected in CO $J=3\rightarrow2$ emission with high significance. The integrated line flux density (''moment 0") maps are shown in Figure~\ref{fig:pstamps} as contours overlaid on the {\it HST}/ACS F814W-band images. 
For each galaxy, we derive spectral profiles by integrating over the full extent of the CO emission.  We define a bounding box for integration by identifying $\ge5\sigma$ pixels in the moment 0 map, and extending the box boundaries to include all pixels within $2.5\sigma_{\mathrm{maj}}$ of the bright pixels, where $\sigma_{\mathrm{maj}}$ ($=\theta_{\mathrm{maj}}/2.35$) is the size of the beam major axis.  Sources with extended emission (identified by eye from the 3D spectral cubes) have their bounding boxes increased to encompass the full line emission. 

\begin{figure*}
\begin{center}
\includegraphics[width=17.5cm]{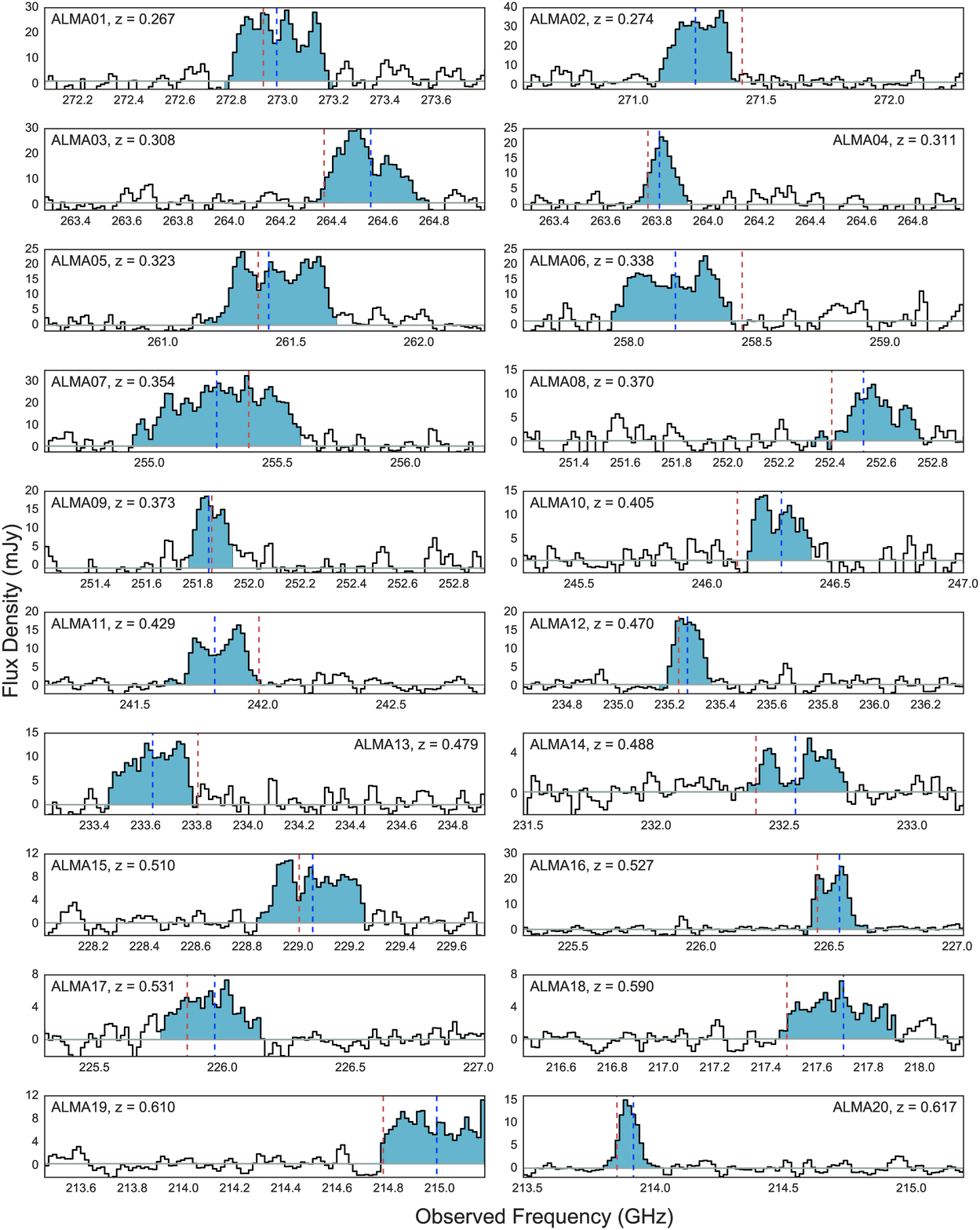}
\caption{Spectral profiles of the CO $J=3\rightarrow2$ emission for the 20 LIRGs measured by ALMA. The source identifications and redshifts are labeled on each panel. For each galaxy, the blue shaded region indicates the channels with line emission, over which we integrate to compute the integrated line flux densities listed in Table~\ref{tab:almaderivedvals}. The dashed vertical lines show the centre of the CO line emission (blue) and the expected locations of the redshifted CO line from the optically determined redshifts (red).}
\label{fig:specprof}
\end{center}
\end{figure*}

Figure~\ref{fig:specprof} shows the CO $J=3\rightarrow2$ spectral profiles for these 20 galaxies. All show clear detections near the optically determined spectroscopic redshifts, with $\sim$10-40 adjacent channels with positive flux densities. Only a few of these galaxies have line profiles that suggest rotation (i.e. one or two Gaussian peaks); most show complex dynamics. A detailed analysis of the galaxy dynamics is beyond the scope of this paper. For this work, we are only interested in the integrated line intensity, $S_{\mathrm{CO(3-2)}}\,\Delta v$, which is proportional to the total molecular gas mass. We measure this by integrating the spectral line profiles over the channels containing the line emission, which we identify by eye from the 3D spectral cubes (highlighted in Figure~\ref{fig:specprof}), and performing continuum subtraction. We list these values of $S_{\mathrm{CO(3-2)}}\,\Delta v$ in Table~\ref{tab:almaderivedvals}, along with the observed centre frequencies of the line emission ($\nu_{\mathrm{CO(3-2),obs}}$, defined as the midway point over all channels with line emission).

The observed CO emission is offset from the optical emission lines by $\approx150$\,km\,s$^{-1}$ on average. This is consistent with the typical errors on the optically determined spectroscopic redshifts ($\sim0.001$), and is typical of velocity differences observed in the optical and CO line emission of star-forming galaxies.\cite[e.g.][]{2002ApJ...580L..21W,2015ApJ...803....6M} 

Since most of these galaxies show complex dynamics, we do not include estimates of line widths in Table~\ref{tab:almaderivedvals}. However, it is interesting to note that many show very broad lines. Considering the equivalent width containing 95\% of the emission, 13 galaxies (65\% of the sample) have lines in excess of 300\,km\,s$^{-1}$, consistent with the broad lines seen in merging systems of galaxies.

We note that adjacent channels, even those far from the line emission, are strongly correlated, with correlation coefficients of $0.65-0.75$. These channel correlations are likely inherent to the ALMA correlator itself and were not removed in the bandpass calibration. The uncertainty on the measured integrated line fluxes will thus be larger than that derived by propagating the calculated rms in the channels maps. As a conservative estimate for the uncertainty in $S_{\mathrm{CO(3-2)}}\,\Delta v$, we compute the rms from the flux density profiles in the channels free of line emission and include a 10\% systematic absolute flux uncertainty, and cite these values in Table~\ref{tab:almaderivedvals}. Similarly, the uncertainties on all quantities that are derived from the integrated line flux densities described in this section are propagated from these values.


We measure CO line luminosities ($L^{\prime}_{\mathrm{CO(3-2)}}$) from the integrated line flux densities ($S_{\mathrm{CO(3-2)}}\,\Delta v$) and observed centre frequencies using Equation 3 from \citet{2005ARA&A..43..677S}.  CO $J=1\rightarrow0$ line luminosities can be estimated from $L^{\prime}_{\mathrm{CO(3-2)}}$ by assuming a CO excitation ladder and conversion factors $R_{J1} = L^{\prime}_{\mathrm{CO(J-(J-1))}} / L^{\prime}_{\mathrm{CO(1-0)}}$. In this work, we adopt a conversion factor of  $R_{31} = 0.625 \pm 0.335$  to convert from $L^{\prime}_{\mathrm{CO(3-2)}}$ to $L^{\prime}_{\mathrm{CO(1-0)}}$. This conversion factor is the median value from a sample of 70 (U)LIRGs in the local universe  \citep{2012MNRAS.426.2601P}, and the uncertainty is propagated through all measurements based on $L^{\prime}_{\mathrm{CO(1-0)}}$. We discuss the possible ramifications of this choice of $R_{31}$ in Appendix \ref{sec:r31}.

\begin{table*}
\centering
\caption{Gas and Dust Properties of $z = 0.25$-0.65 LIRGs}
\label{tab:almaderivedvals}
\begin{tabular}{cccccccc} 
\hline
\hline
ALMA ID & $S_{\mathrm{1 mm}}$$^\mathrm{a}$ & $\nu_{\mathrm{CO(3-2),obs}}$ & $S_{\mathrm{CO(3-2)}}\,\Delta v$ & $L^{\prime}_{\mathrm{CO(3-2)}}$ & log($M_{\mathrm{H}_2}$)$^\mathrm{b}$ & $f_{\mathrm{H}_2}$$^\mathrm{b}$ & SFE$^\mathrm{b}$ \\
& mJy & GHz & Jy\,km\,s$^{-1}$ & $10^9$\,K\,km\,s$^{-1}$\,pc$^2$ & log$(M_{\odot})$  & $M_{\mathrm{H}_2} / M_{*}$ & Gyr$^{-1}$ \\
\hline
ALMA01 & $<1.09$ & 272.977 & $ 7.74 \pm 0.85$ & $ 3.26 \pm 0.36$ & $10.30 \pm 0.10$ & $ 0.71 \pm 0.18$ & $ 0.49 \pm 0.12$\\ 
ALMA02 & $1.26 \pm 0.32$ & 271.243 & $ 8.03 \pm 0.84$ & $ 3.56 \pm 0.37$ & $10.35 \pm 0.10$ & $ 1.00 \pm 0.25$ & $ 0.88 \pm 0.21$\\ 
ALMA03 & $0.73 \pm 0.18$ & 264.552 & $ 7.01 \pm 0.76$ & $ 3.96 \pm 0.43$ & $10.39 \pm 0.10$ & $ 0.99 \pm 0.24$ & $ 0.80 \pm 0.19$\\ 
ALMA04 & $<0.36$ & 263.810 & $ 2.57 \pm 0.31$ & $ 1.48 \pm 0.18$ & $ 9.95 \pm 0.10$ & $ 0.20 \pm 0.05$ & $ 1.13 \pm 0.28$\\ 
ALMA05 & $<0.57$ & 261.414 & $ 8.54 \pm 0.89$ & $ 5.33 \pm 0.56$ & $10.51 \pm 0.10$ & $ 0.45 \pm 0.11$ & $ 0.75 \pm 0.18$\\ 
ALMA06 & $1.03 \pm 0.32$ & 258.182 & $ 6.59 \pm 0.76$ & $ 4.52 \pm 0.52$ & $10.44 \pm 0.10$ & $ 0.21 \pm 0.05$ & $ 0.76 \pm 0.19$\\ 
ALMA07 & $<0.35$ & 255.263 & $14.86 \pm 1.55$ & $11.22 \pm 1.17$ & $10.83 \pm 0.10$ & $ 0.56 \pm 0.13$ & $ 0.36 \pm 0.09$\\ 
ALMA08 & $<0.34$ & 252.530 & $ 2.49 \pm 0.35$ & $ 2.06 \pm 0.29$ & $10.14 \pm 0.11$ & $ 1.07 \pm 0.31$ & $ 0.91 \pm 0.33$\\ 
ALMA09 & $<0.28$ & 251.842 & $ 2.84 \pm 0.35$ & $ 2.39 \pm 0.30$ & $10.17 \pm 0.11$ & $ 0.45 \pm 0.12$ & $ 1.07 \pm 0.27$\\ 
ALMA10 & $<0.33$ & 246.290 & $ 2.59 \pm 0.31$ & $ 2.58 \pm 0.31$ & $10.21 \pm 0.10$ & $ 0.45 \pm 0.12$ & $ 1.04 \pm 0.52$\\ 
ALMA11 & $<0.30$ & 241.812 & $ 3.47 \pm 0.38$ & $ 3.90 \pm 0.43$ & $10.38 \pm 0.10$ & $ 0.61 \pm 0.15$ & $ 0.67 \pm 0.19$\\ 
ALMA12 & $<0.28$ & 235.270 & $ 2.63 \pm 0.29$ & $ 3.56 \pm 0.39$ & $10.36 \pm 0.10$ & $ 0.82 \pm 0.20$ & $ 1.20 \pm 0.29$\\ 
ALMA13 & $<0.51$ & 233.627 & $ 3.78 \pm 0.43$ & $ 5.33 \pm 0.61$ & $10.52 \pm 0.10$ & $ 0.85 \pm 0.21$ & $ 0.97 \pm 0.24$\\ 
ALMA14 & $<0.21$ & 232.544 & $ 1.19 \pm 0.15$ & $ 1.75 \pm 0.22$ & $10.05 \pm 0.11$ & $ 0.35 \pm 0.09$ & $ 1.82 \pm 0.47$\\ 
ALMA15 & $<0.41$ & 229.057 & $ 3.63 \pm 0.39$ & $ 5.83 \pm 0.63$ & $10.56 \pm 0.10$ & $ 0.81 \pm 0.20$ & $ 0.64 \pm 0.16$\\ 
ALMA16 & $0.53 \pm 0.13$ & 226.540 & $ 3.89 \pm 0.41$ & $ 6.69 \pm 0.70$ & $10.63 \pm 0.10$ & $ 1.07 \pm 0.26$ & $ 0.43 \pm 0.14$\\ 
ALMA17 & $0.27 \pm 0.07$ & 225.970 & $ 2.09 \pm 0.24$ & $ 3.65 \pm 0.41$ & $10.36 \pm 0.10$ & $ 0.51 \pm 0.13$ & $ 1.01 \pm 0.25$\\ 
ALMA18 & $0.18 \pm 0.04$ & 217.703 & $ 2.27 \pm 0.26$ & $ 4.91 \pm 0.55$ & $10.48 \pm 0.10$ & $ 0.46 \pm 0.12$ & $ 0.92 \pm 0.22$\\ 
ALMA19 & $0.55 \pm 0.11$ & 214.989 & $ 3.63 \pm 0.39$ & $ 8.42 \pm 0.90$ & $10.81 \pm 0.10$ & $ 5.80 \pm 1.47$ & $ 0.58 \pm 0.14$\\ 
ALMA20 & $<0.10$ & 213.915 & $ 1.80 \pm 0.20$ & $ 4.27 \pm 0.47$ & $10.43 \pm 0.10$ & $ 0.55 \pm 0.14$ & $ 0.82 \pm 0.23$\\

\hline
\end{tabular}
\begin{flushleft}
$^{\mathrm{a}}$1\,mm continuum flux density measurements (Section~\ref{ssec:almaimage}), with 3$\sigma$ upper limits provided for non-detections. \\
$^{\mathrm{b}}$See Section~\ref{sec:phys} for estimates of the gas properties.\\
\end{flushleft}
\end{table*}

\section{Multi-wavelength Data}\label{sec:data}

\subsection{COSMOS multiwavelength data}\label{sec:cosmos}

The COSMOS field offers deep, multi-wavelength coverage over a wide 2 deg$^{2}$ area \citep{2007ApJS..172....1S}. Especially important for our study are new, deep near-IR data from UltraVISTA \citep{2012A&A...544A.156M} and the {\em Spitzer} Large Area Survey with Hyper-Suprime-Cam (SPLASH; P.\ Capak et al.\ 2016, in prep) project that allow us to measure accurate stellar masses \citep{2016arXiv160402350L}. The stellar masses are derived from the stellar population synthesis models of \citet{2003MNRAS.344.1000B} applied to the full 30-band UV/optical/near-IR data available in COSMOS.

In addition, COSMOS has deep far-infrared imaging with {\em Herschel} PACS and SPIRE from 100--500 $\mu$m \citep{2013ApJ...778..131L}. Detections in multiple {\em Herschel} bands are necessary for accurately fitting the full infrared SED and measuring accurate SFRs, especially in starburst galaxies \citep{2015ApJ...801...80L}. All of the LIRGs in our sample were detected in multiple {\em Herschel} bands, and we have updated SPIRE fluxes using a new prior-based source extraction tool based on the Bayesian inference tool Stan \citep[XID$+$;][]{2017MNRAS.464..885H}.

\begin{figure*}
\centering
\includegraphics[width=\textwidth]{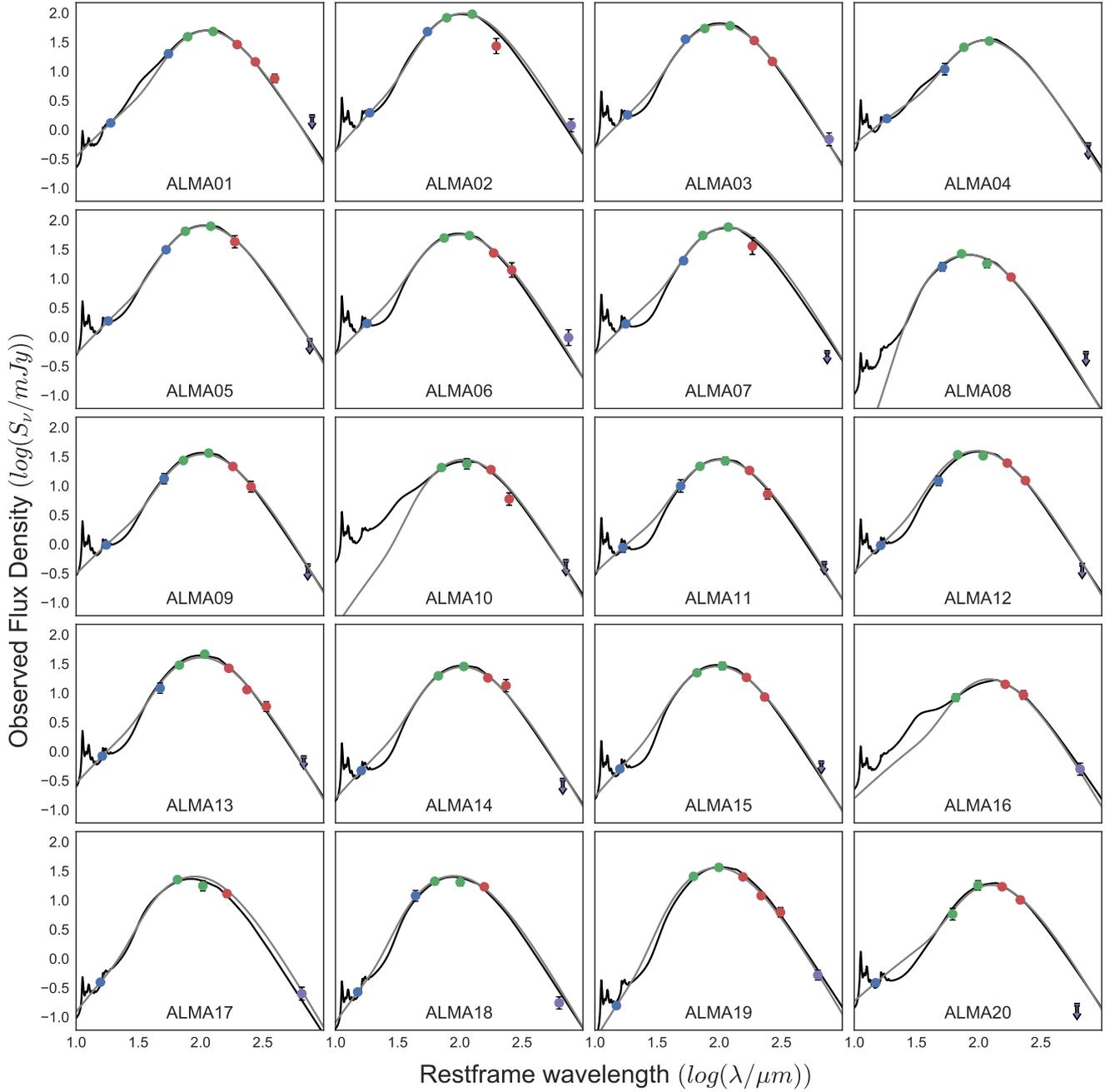} 
\caption{Infrared Spectral Energy Distributions (SEDs) of the galaxies in our sample. Circles represent the observed flux densities from {\em Spitzer} (blue; 24 $\mu$m and 70 $\mu$m), {\em Herschel} PACS (green; 100 $\mu$m and 160 $\mu$m), ({\em Herschel} SPIRE (red; 250 $\mu$m, 350 $\mu$m, and 500 $\mu$m), and ALMA (purple; 1 mm).  All observed flux densities also have error bars plotted in black, although many of these error bars are smaller than the size of the plotted circles and are not visible.  Sources without a detection at ALMA 1 mm have 5$\sigma$ upper limits plotted as downward arrows. Solid lines represent the best fit SEDs from fitting to \citet{2007ApJ...657..810D} dust models (black) and \citet{2012MNRAS.425.3094C}  greybody plus power-law models (gray). }
\label{fig:seds}
\end{figure*}

We fit the full infrared photometry (including 1 mm continuum from ALMA when detected) to the physically motivated \citet{2007ApJ...657..810D} dust models. These SED fits provide estimates of the dust mass, dust temperature, and total infrared luminosity for the galaxies in our sample. We note that we have also performed the full analysis in this paper using infrared properties derived from fitting greybody plus power-law models from \citet{2012MNRAS.425.3094C}, and do not find significant differences in our results. Figure \ref{fig:seds} displays the full far-infrared SEDs along with the model fits from \citet{2007ApJ...657..810D} and \citet{2012MNRAS.425.3094C}.  The SED fits from both methods agree quite well when provided with multiple far-infrared photometric points. We note that the ALMA 1 mm continuum $5\sigma$ upper limit in ALMA20 lies significantly below the predicted 1 mm flux from the SED fits.  It is unclear why this continuum upper limit is so low, as the CO line is detected with high significance (Figure \ref{fig:specprof}), and the far-infrared photometry is unlikely to be confused  (Figure \ref{fig:irac_cutouts}). Despite this mismatch in the continuum photometry, we include this source in the remaining analysis, but note that excluding the source does not have a significant impact on the results.  

\subsection{Optical lines}

A spectroscopic survey of the COSMOS field has recently been completed with Hectospec on MMT \citep{2005PASP..117.1411F}. The survey is $\geq90\%$ complete to a limiting magnitude of $r>20.8$ \citep{2015ApJ...815..104D,2016ApJ...821..101Z} and provides optical spectroscopic redshifts for $\sim6000$ galaxies in the COSMOS field at $0\leq z \leq 0.6$. A large fraction of the data is of high enough quality for detailed emission line analysis. These observations provide accurate metallicities for an unbiased sample of galaxies and allow us to identify AGN.  

12 of the 20 LIRGs in our sample were observed in the Hectospec campaign.  Emission line fluxes are measured as described in \citet{2011ApJ...730..137Z}.  8 sources are identified as AGN from their location on the [NII]/H$\alpha$ vs [OIII]/H$\beta$ diagram, using the \citet{2006MNRAS.372..961K} classification. These AGN are largely missed by other AGN diagnostics, as none of them have photometry that meet the IRAC power-law criteria that identifies obscured AGN from reprocessed radiation at mid-infrared wavelengths \citep{2008ApJ...687..111D,2012ApJ...748..142D}, and only one of these sources was detected in the X-ray by the {\em Chandra} observations of COSMOS \citep[C-COSMOS;][]{2009ApJS..184..158E,2016ApJ...819...62C}. 

For the remaining 4 sources with Hectospec observations, accurate gas-phase metallicities can be determined from the well established strong-line method \citep{2004ApJ...617..240K}. These metallicities will allow us to more accurately convert CO line luminosity to gas mass (see Appendix \ref{sec:alpha_co}).

\subsection{Complimentary Data}

In order to expand the sample size for our study, we have compiled the majority of the published galaxies with observations of both $\lco$ and $\lir$. An analysis of this full population is presented in \citet{Cortzen_Thesis}. In this paper, we compare our results to the subset of galaxies in the literature at a similar redshift range to our sample \citep[$z \sim 0.1$--0.6;][]{2011A&A...528A.124C,2013ApJ...768..132B,2014ApJ...796...63M} in order to avoid any effects from possible redshift evolution.  While these studies all contain observations of $\lco$, they were all selected with different criteria and are subject to unique selection biases. Thus, we include them for a better perspective on the full dynamic range of the overall galaxy population, but do not include them in any fits to the data or detailed, quantitative analysis. 

The studies with published observations of $\lir$ and $\lco$ with redshifts $z \sim 0.1$--0.6 are:
\begin{itemize}
\item \citet{2011A&A...528A.124C} - CO observations of 36 ULIRGs with log($L_{\rm{FIR}}) > 12.45$ at $z = 0.2$--0.6 originally detected at 60 $\mu$m (IRAS, ISO). Infrared luminosities are calculated from IRAS fluxes as $L_{\rm{FIR}} = 1.26 \times 10^{-14} (2.58\ S_{60} + S_{100})$ W\,m$^{-2}$ \citep{1996ARA&A..34..749S} and assuming $L_{\rm{IR}} = 1.3 \times L_{\rm{FIR}}$. Stellar masses were provided by F. Combes (private communication).
\item \citet{2013ApJ...768..132B} - CO observations of 31 SFGs at $z = 0$--0.5 with CARMA (CO $J=1\rightarrow0$ at $z < 0.3$ and CO $J=3\rightarrow2$ at $z > 0.3$).  Sources are drawn from the Sloan Digital Sky Survey Data Release 7, and star formation rates and stellar masses are calculated from the SDSS emission lines and photometry, respectively. We convert star formation rates to infrared luminosity as $\lir / L_{\odot} = SFR [M_{*} / \mathrm{yr}] / (8.6 \times 10^{-11} ) $, as in \citet{2013A&A...558A..67A}. This conversion relies on an accurate accounting for dust obscuration in the derivation of the SFRs from the SDSS emission lines and assumes that the bulk of the emission produced by the star formation will be absorbed by dust and reradiated in the far-infrared, which is likely true at these elevated SFRs \citep{2015ApJ...801...80L}. 
\item \citet{2014ApJ...796...63M} - CO observations of 15 galaxies at $z = 0.219$--0.887 with {\em Herschel} $S_{250} > 150$ mJy. Infrared luminosities are determined from fitting the full infrared photometry from {\em Spitzer} and {\em Herschel} to \citet{2007ApJ...657..810D} dust models. Stellar masses are not available for these galaxies. 
\end{itemize}

\subsection{Distance to the Main Sequence}\label{sec:ms}

The main sequence of star forming galaxies describes the typical star formation rate in a galaxy at a given stellar mass and redshift. Commonly, the main sequence is derived by binning star forming galaxies by stellar mass, measuring the median SFR in each stellar mass bin, and then finding the best fit-relationship to describe the median SFRs (typically a power-law). The standard deviation of SFRs in each stellar mass bin is used to describe the width of the distribution, and this width has been found to be a fairly constant $\sigma \sim 0.3$ dex, independent of both stellar mass and redshift. The phrase ``main sequence'' is often used to describe both the best-fit relationship between median SFRs, and to describe the entire population of galaxies that lie within the typical width of the distribution of SFRs around the median. To avoid confusion, in this paper we will use the phrase ``main sequence'' to describe the median SFR expected for a given stellar mass and redshift, and when referring to the distribution of SFRs around the median, we will use the phrase ``main sequence distribution''. 

For every galaxy, we can define the ``distance to the main sequence'' as the ratio between its measured star formation rate and the expected (median) star formation rate of a star forming galaxy at the same stellar mass and redshift. In log units, we define the distance to the main sequence as $\Delta SFR_{MS} \equiv log(SFR/SFR_{MS})$. In the literature, this is sometimes described in terms of specific star formation rate ($SSFR \equiv SFR/M_{*}$), but since these are ratios of galaxies with the same stellar mass, the distance to the main sequence is identical when dealing with $SFR$ or $SSFR$. $\Delta SFR_{MS} $ is a measure of how unusual a galaxy's $SFR$ is for its stellar mass and redshift, with the majority of galaxies expected to lie within $\pm 0.3$ dex of the main sequence \citep{2007ApJ...660L..43N}.

The main sequence has been studied in detail for about a decade, and while its existence is well-established, there is little consensus in the slope or normalization of the relationship. \citet{2014ApJS..214...15S} compiled much of the existing literature and were able to find some consensus by applying common calibrations. The time-dependent best-fit relationship from their compilation was $SFR_{MS}(M_{*}, t) =  (0.84 - 0.026 \times t) log(M_{*}) - (6.51 - 0.11 \times t)$. However, recent studies have found that the main sequence is actually more complicated than a single power law, and that it in fact has a stellar mass dependent slope that flattens at large stellar masses \citep{2012ApJ...754L..29W,2014ApJ...795..104W,2015ApJ...801...80L,2016ApJ...817..118T}.  In light of these results, we use the definition of the main sequence provided in Equation 9 of \citet{2015A&A...575A..74S}, who use stacking of {\em Herschel} data to derive a redshift dependent main sequence that includes the flattening of the high stellar mass end.  The stellar mass and redshift dependent main-sequence SFR is defined as:
\begin{equation*}
\mathrm{log}(SFR_{MS}) = m - m_{0} + a_{0}r - a_{1}[\mathrm{max}(0, m - m_{1} - a_{2}r)]^{2}
\end{equation*} 
where $m = \mathrm{log}(M_{*}/10^{9} M_{\odot})$, $r = \mathrm{log}(1+z)$, and $m_{0} = 0.5$, $m_{1} = 0.36$, $a_{0} = 1.5$, $a_{1} = 0.3$ and $a_{2} = 2.5$ \citep{2015A&A...575A..74S}. This relationship is derived using stellar masses and SFRs calculated from a Salpeter IMF, and we use conversion factors of $\mathrm{log}(M_{*, \mathrm{Chab}}) = \mathrm{log}(M_{*, \mathrm{Sal}}) - 0.24$ \citep{2013MNRAS.435...87M} and $SFR_{\mathrm{Chab}} = SFR_{\mathrm{Sal}} / 1.7$ \citep{2012ApJ...757...54Z} to convert between the two IMFs. We apply this main sequence when deriving $\Delta SFR_{MS} $ for the galaxies in our sample and from the literature. We note that for the limited dynamic range in redshift and stellar mass of our galaxies, the particular assumed shape of the main-sequence does not have a significant affect on $\Delta SFR_{MS}$, and adoption of a single power-law MS \citep[e.g. from][]{2014ApJS..214...15S} does not result in significant changes to our results.

\subsection{Statistical Tests and Curve Fitting}

In this paper, we often examine possible correlation between two independently measured variables, and when we find correlation, we wish to find best-fit curves to describe the relationship between the two variables. In most cases, both of the variables in questions have associated measurement uncertainty that must be taken into account. Here we describe the processes by which we perform these tests and fits. 

Given a set of paired measurements ($X_{i}, Y_{i}$), we wish to determine if there is an underlying correlation between $X$ and $Y$. We measure the correlation strength by applying a Spearman Rank correlation test \citep{1904AJP...15...72} as implemented in the SciPy package \citep{2001SciPy}  for Python\footnote{\url{http://docs.scipy.org/doc/scipy-0.14.0/reference/generated/scipy.stats.spearmanr.html}}. This nonparametric rank test returns a correlation coefficient ($\rho$) that is interpreted as $\rho = 0$ means no correlation, while $\rho = +1$ or $\rho = -1$ mean perfect increasing or decreasing monotonic correlation, respectively. 

To determine the significance of the correlation, we perform a permutation test, which tests the probability of the null hypothesis. The null hypothesis is that there is no correlation between the two variables $X$ and $Y$, which would mean any of the $Y_{i}$ observations are just as likely to appear with any of the $X$'s, so $Y_{i}$ is just as likely to appear with $X_{i}$ as it is to appear with $X_{j}, i \neq j$.  We test the likelihood that our measured value of $\rho_{\mathrm{meas}}$ was produced given the null hypothesis by permuting the $Y$'s among the $X$'s and recalculating the spearman rank correlation for each permutation ($\rho_{\mathrm{perm}}$). In an ideal scenario we would sample the full $n!$ possible permutations, but often it is enough to probe a large enough random set of permutations. In this paper, we sample $N_{perm} = 10^{6}$ permutations. Then, the probability that the measured $\rho_{\mathrm{meas}}$ was produced under the null hypothesis is simply the fraction of permutations where $\rho_{\mathrm{perm}}$ is larger than $\rho_{\mathrm{meas}}$, or $p = N(\rho_{\mathrm{perm}} > \rho_{\mathrm{meas}}) / N_{perm}$. A $p$-value of $p \leq 0.05 $ is traditionally required to reject the null hypothesis.


When we find statistically significant correlation, we often wish to find the best-fit curves to describe the correlation. In order to account for measurement errors in both $X$ and $Y$ measurements, we use a Bayesian method of linear regression described in \citet{2007ApJ...665.1489K}.  The \citet{2007ApJ...665.1489K} method uses a Markov Chain Monte Carlo (MCMC) algorithm (specifically a Gibbs sampler) to sample the posterior distribution, and it can account for measurement uncertainty in both variables as well as include intrinsic scatter. With the full posterior distribution, we are able to report both the expected parameters and a confidence interval. 

\section{Observed Luminosities}\label{sec:lir_lco}

Star formation efficiencies can be measured by comparing the current SFR (calculated from $\lir$) to the gas mass (calculated from $\lco$). However, before making this comparison, it is instructive to directly compare the observed properties $\lir$ and $\lco$ before any uncertain extrapolations and conversions are applied. 

\begin{figure*}
\centering
\includegraphics[width=\textwidth]{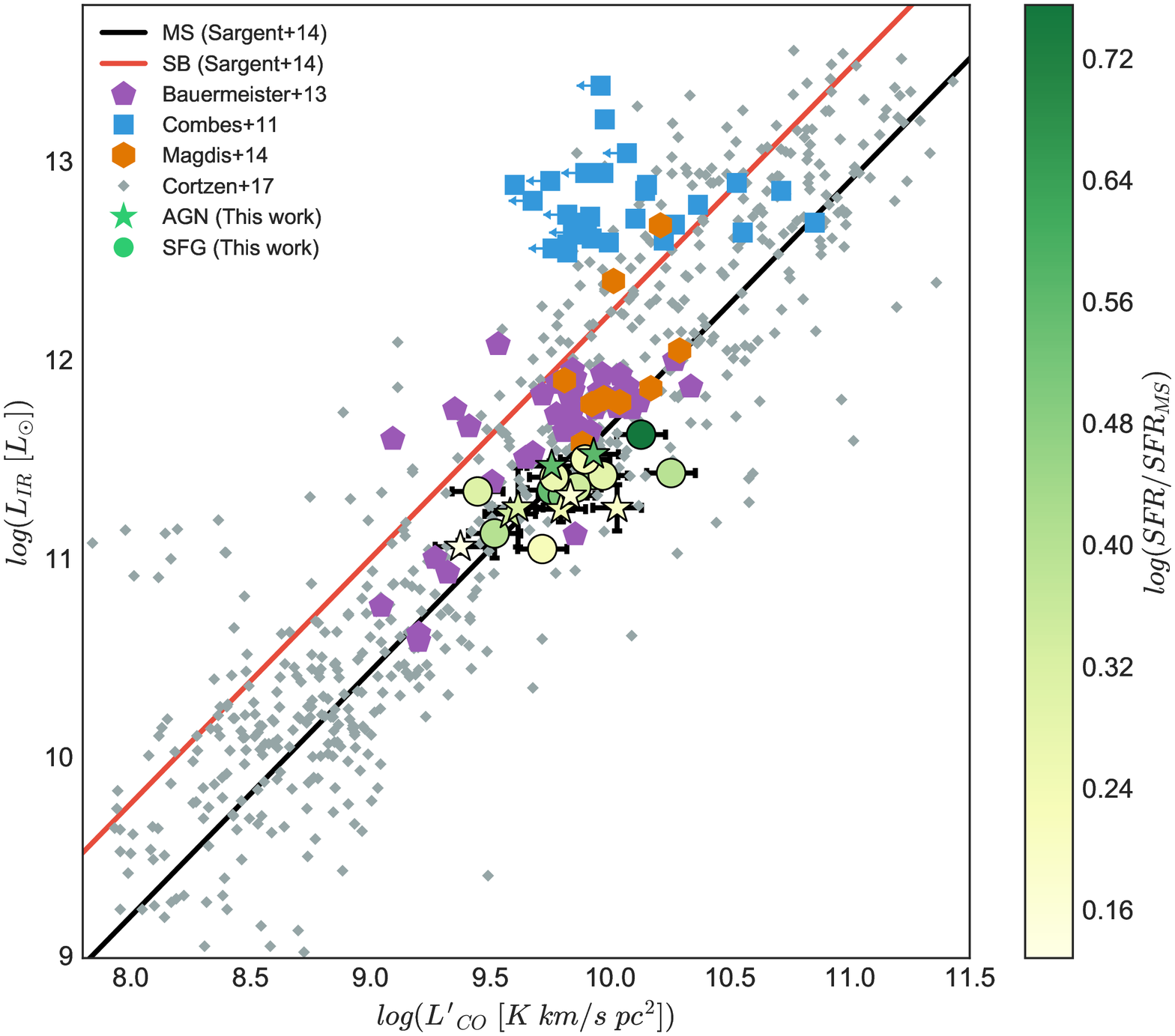}
\caption{Infrared luminosities and CO(1-0) line luminosities of galaxies in our sample, along with many of the existing samples published in the literature. Grey points represent the bulk of published studies compiled in Cortzen et al. (in prep). Blue \citep{2011A&A...528A.124C}, purple \citep{2013ApJ...768..132B}, and orange \citep{2014ApJ...796...63M} symbols represent the galaxies with published $\lir$ and $\lco$ observations at $z \sim 0.1$--0.6. Black and red lines represent observed relationships from the literature for main sequence and starburst galaxies, respectively \citep{2014ApJ...793...19S}.  The galaxies in our sample are plotted as circles (SFGs) and stars (AGN), with colour indicating distance to the main sequence, as indicated by the vertical colour bar. }
\label{fig:lir_lco}
\end{figure*}

In Figure \ref{fig:lir_lco}, we plot $\lir$ and $\lco$ of our 20 LIRGs, along with numerous values from the literature. The grey background of sources contains much of the published $\lir$ and $\lco$ values in the literature \citep[][and references within]{Cortzen_Thesis}. 
The blue, purple, and orange symbols represent the galaxies with published $\lir$ and $\lco$ observations at a similar redshift range to our study \citep[$z \sim 0.1$--0.6;][]{2011A&A...528A.124C,2013ApJ...768..132B,2014ApJ...796...63M}. Our sample of 20 LIRGs is plotted in green colours, and galaxies containing an AGN (plotted as stars) are marked differently from pure star forming galaxies (SFGs; plotted as circles).  The shade of green in each symbol indicates its distance to the main sequence, as indicated by the vertical colour bar.  Black and red lines represent expected relationships for main sequence and starburst galaxies, respectively \citep{2014ApJ...793...19S}. 

The LIRGs in our sample mostly scatter around the track for main sequence galaxies. Location in the $\lir$-$\lco$ plane does not seem to depend on distance to the star-forming main sequence. This matches well with the other galaxies in this redshift range with $\lir \lesssim 10^{12} L_{\odot}$. At this redshift, it is only ULIRGs ($\lir \geq 10^{12} L_{\odot}$) that appear to lie on or above the starburst sequence from \citet{2014ApJ...793...19S}. 

At a fixed $\lco$, AGN host galaxies may be expected to have a higher $\lir$, either because the AGN is contributing extra dust heating that increases the $\lir$, or because of the presence of a coeval burst of star formation that coincides with the growth of the supermassive black hole \citep[e.g.][]{2013ARA&A..51..105C}. However, the AGN in our sample do not scatter apart from the pure SFGs, and indeed lie very close to the ``main sequence'' line from \citet{2014ApJ...793...19S}.  The lack of a strong mid-IR power-law in these galaxies suggests that the AGN is not an important source of heating for the dust, and may explain why these galaxies do not have an elevated $\lir / \lco$ ratio, despite the presence of an AGN. Another explanation may be that the AGN in our sample have different CO excitation, as we discuss in more detail in Appendix \ref{sec:r31}. 

In Figure \ref{fig:lir_lco_ratio}a, we examine how the $\lir / \lco$ ratio explicitly depends on the distance to the main sequence, again including all available galaxy samples at $z \sim 0.1$--0.6. Focusing only on the galaxies in our sample, there does not appear to be a clear trend with distance to the main sequence. 

\begin{figure*}
\centering
\includegraphics[width=\textwidth]{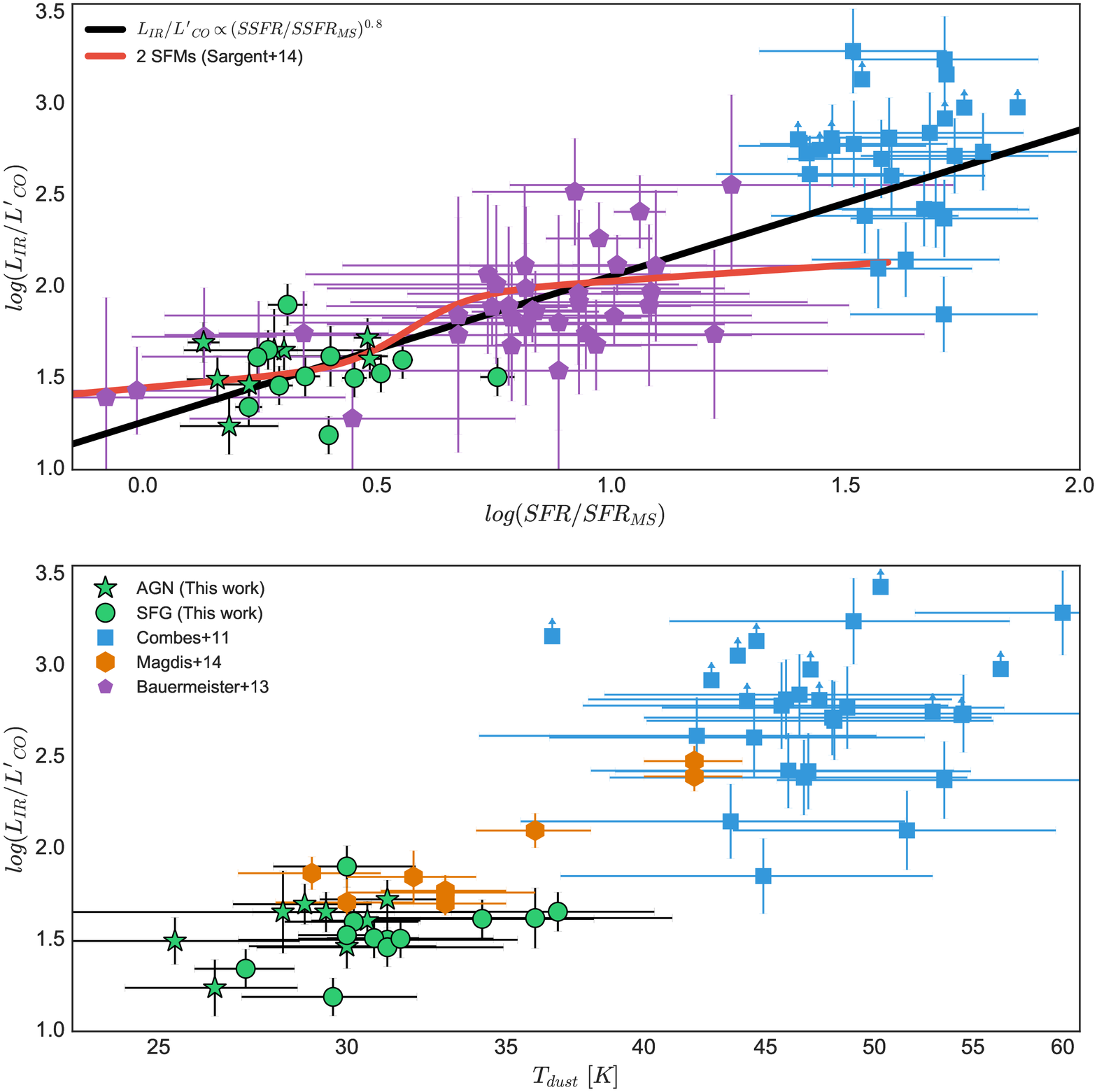}
\caption{$\lir / \lco$ ratio as a function of distance from the main sequence (top) and $T_{dust}$ (bottom). Symbols are the same as in Figure \ref{fig:lir_lco}, with blue \citep{2011A&A...528A.124C}, purple \citep{2013ApJ...768..132B}, and orange \citep{2014ApJ...796...63M} symbols representing galaxies from the literature.  Green circles (SFGs) and stars (AGN) are galaxies in our sample. In the top panel, the solid lines represent possible evolutionary tracks of $\lir/\lco$ representing smooth evolution \citep[black;][]{2012ApJ...760....6M} and step-like evolution \citep[red][]{2014ApJ...793...19S}.  The track from \citet{2014ApJ...793...19S} is for a galaxy with stellar mass log($M_{*}) = 10.7$, scaled to the data. }
\label{fig:lir_lco_ratio}
\end{figure*}

\citet{2012ApJ...760....6M} studied the dependence of the $\lir / \lco$ ratio on distance to the main sequence in galaxies at $0.5 < z < 2$ and found that the relationship could be described by either a single smooth relationship ($\lir / \lco \propto (SSFR/SSFR_{MS})^{0.8})$ or a more step-like sequence representative of two star formation modes, as described in \citet{2014ApJ...793...19S}. Both of these possible tracks are plotted in Figure \ref{fig:lir_lco_ratio}a, but unfortunately our observations alone cannot distinguish between the two possibilities.

If we include sources from the other studies at $z \sim 0.1$--0.6, we see evidence of increasing $\lir / \lco$ ratio with distance to the main sequence, with the galaxies at high $\Delta SFR_{MS}$ favouring the smooth evolution proposed by \citet{2014ApJ...793...19S}. However, much of this evolution is driven almost entirely by the ULIRGs from \citet[][in blue]{2011A&A...528A.124C}. As discussed previously, combining observations from disparate studies to measure evolution can be misleading due to the very different selection functions of these studies. Figure \ref{fig:lir_lco_ratio}b shows the available dust temperatures of all the galaxies at $z \sim 0.1$--0.6. Our galaxies and the \citet{2012ApJ...760....6M} sample were selected at 250 $\mu$m and are biased toward cold dust temperatures $T_{d} \lesssim 40$ K while the \citet{2011A&A...528A.124C} galaxies were selected at 60 $\mu$m and are biased toward hot dust temperatures $T_{d} \gtrsim 40$ K. Because of these severe selection effects, it is not clear whether the large distance to the main sequence or the hot dust temperatures are driving the elevated $\lir / \lco$ ratio in the \citet{2011A&A...528A.124C} sample. \citet{2014A&A...561A..86M} find a correlation between dust temperature and distance to the main sequence, which suggests that these might not be separate selection effects, but simply an accurate reflection of the galaxies that populate different parts of the main sequence. However, the correlations found there only extend to dust temperatures $T_{d} \lesssim 40$ K and they find only moderate temperature evolution of $\sim 10$ K on and off the main sequence. In addition, dust temperature is also known to correlate with infrared luminosity \citep[e.g.][]{2013ApJ...778..131L}, which may also explain the extremely high temperatures of the \citet{2011A&A...528A.124C} sources. 

\vspace{0.5em}
The observed $\lir$ and $\lco$ from our sample show little dependence on distance to the main sequence. When taken together with the other published CO studies at a similar redshift range, we see evidence for an increasing $\lir / \lco$ ratio with $\Delta SFR_{MS}$. However, it is not clear if this increasing ratio is truly driven by distance to the main sequence, dust temperature, or some other effect or bias because the increase is almost entirely due to the inclusion of a single study with a very different selection function.

\section{Morphology}\label{sec:morph}

Galaxy morphologies provide important insight into the environments in which star formation is taking place. In particular, studies of infrared luminous galaxies in the local universe demonstrate a high incidence of major mergers, and even find a correlation between infrared luminosity and merger stage \citep{2002ApJS..143..315V,2004PhDT........18I}. In the local universe, the majority of galaxies with $L_{\rm{IR}} \lesssim 10^{11.5} L_{\odot}$ have disk morphologies while higher luminosity galaxies are increasingly powered by major mergers \citep[e.g.][]{2010ApJ...721...98K,2011AJ....141..100H,2016ApJ...825..128L}.

While the connection between mergers and (U)LIRGs is well-established in the local universe \citep[e.g.][]{1996ARA&A..34..749S}, it is less clear what role mergers play in infrared luminous galaxies at high redshift. (U)LIRGs are much more common in the high redshift universe and have a significant contribution to the buildup of stellar mass and growth of supermassive black holes, producing as much as 50\% of the stellar mass in the universe at redshifts $z \sim 2$--3 \citep{2005ApJ...622..772C,2005ApJ...632..169L,2012ApJ...761..140C} and as much as $\sim 30$\% of the integrated black hole growth through highly obscured accretion \citep{2009ApJ...706..535T,2010ApJ...722L.238T}. However, the rising normalization of the star-forming main sequence (Section \ref{sec:ms}) implies that (U)LIRGs at high redshift are no longer extreme starburst galaxies, but are rather more similar to normal, main sequence galaxies and may not require major mergers to power their luminosities. 

\begin{figure*}
\centering
\includegraphics[width=\textwidth, trim={100, 70, 100, 30}]{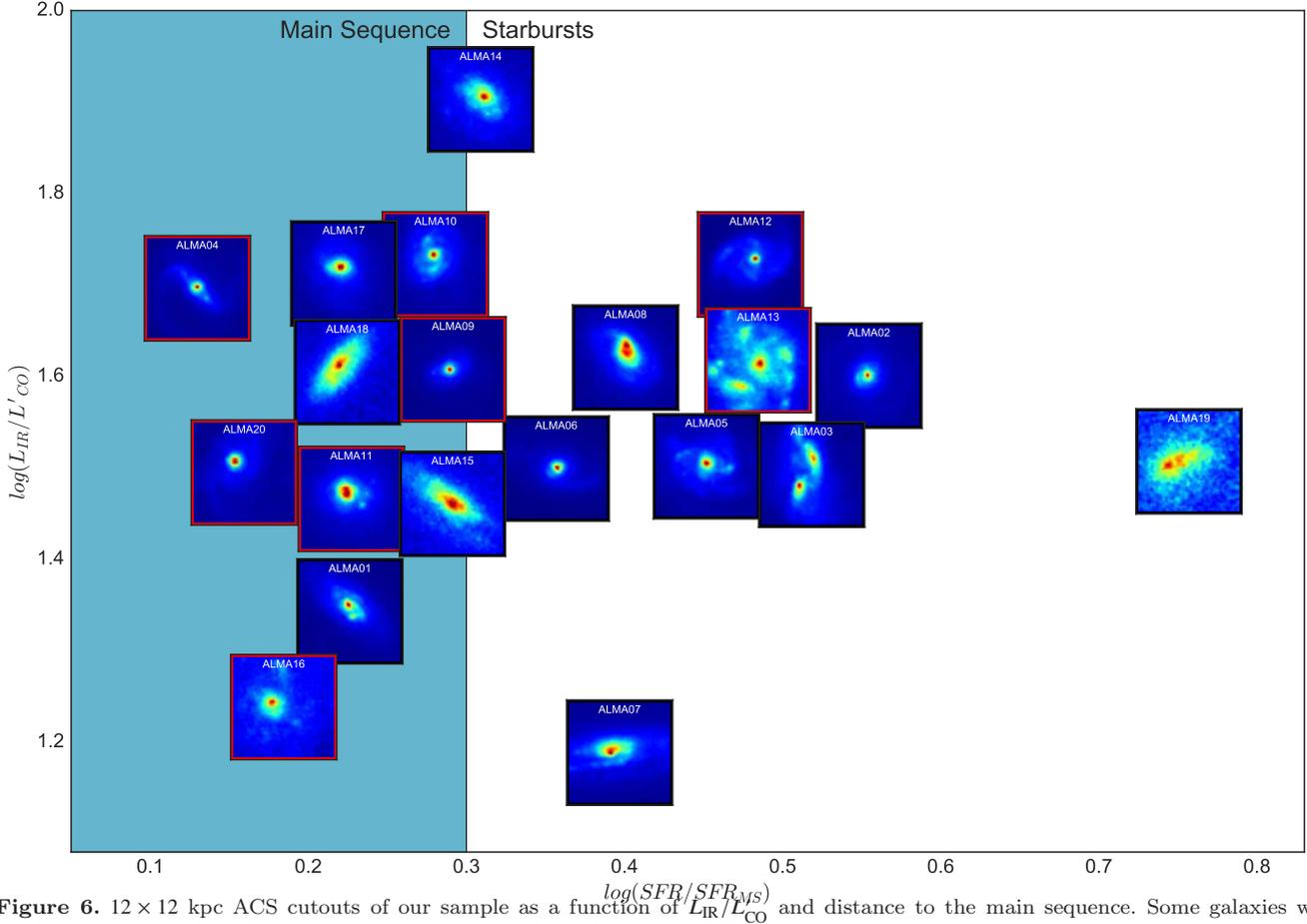}
\caption{$12 \times 12$ kpc ACS cutouts of our sample as a function of $\lir / \lco$ and distance to the main sequence. Some galaxies with overlapping cutouts have had their positions slightly adjusted for clarity. The vertical line and background shading represents the typical distinction between main sequence and starburst galaxies.  Galaxies identified as AGN have red borders while pure star-forming galaxies have black borders.}
\label{fig:cutouts}
\end{figure*}

In Figure \ref{fig:cutouts}, we plot restframe optical cutouts centered on each of the galaxies in our sample, as observed by the Advanced Camera for Surveys (ACS) instrument on the Hubble Space Telescope (HST). Each cutout has been scaled to a physical size of $12 \times 12$ kpc. The cutouts are plotted as a function of their $\lir / \lco$ ratio (a proxy for star formation efficiency) and distance to the main sequence. Galaxies identified as containing AGN have red borders while pure star forming galaxies have black borders. We do not see any systematic morphological differences between AGN and star-forming galaxies. 

The galaxies in our sample have predominantly disk-like morphologies, with only two showing irregular features that suggest an ongoing or past merger (ALMA03 and ALMA19). ALMA13 also shows a very clumpy disk structure, which may be due to merging or some other process that can disrupt galaxy structure such as disk instabilities.  All of these galaxies with irregular features have a slightly elevated distance to the main sequence, but do not seem to be outliers in $\lir / \lco$. This suggests that mergers may not trigger significantly elevated star formation efficiencies, or that they only trigger a short period of enhanced star formation efficiency that lasts for only a small fraction of the merger process. 

\begin{figure*}
\centering
\includegraphics[width=\textwidth]{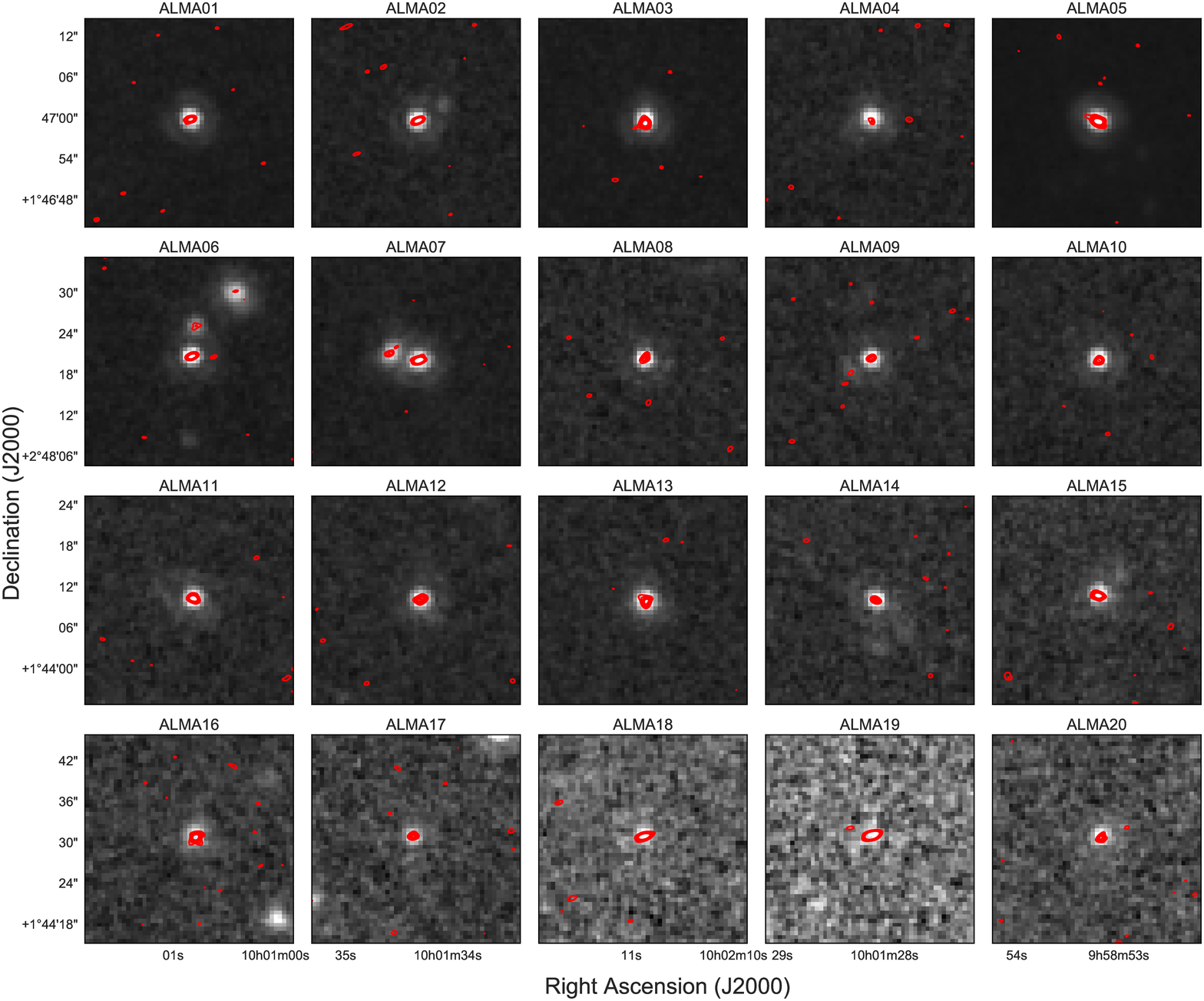}
\caption{$30'' \times 30''$ cutouts from {\em Spitzer} IRAC 8 $\mu$m of our full sample. Red contours represent CO(3-2) emission detected at 3$\sigma$ or higher. ALMA06 and ALMA07 have infrared luminous companions that are at the same redshift (based on both photometric redshifts and the marginal detection of CO(3-2)). The cutout size represents the approximate size of the {\em Herschel} SPIRE 250 $\mu$m beam, which suggests that the far-infrared fluxes in ALMA06 and ALMA07 are likely blended.}
\label{fig:irac_cutouts}
\end{figure*}

In Figure \ref{fig:irac_cutouts}, we plot IRAC 8 $\mu$m cutouts of our full sample, with CO(3-2) contours overplotted in red. Each of these cutouts is $30'' \times 30''$, which is similar to the beamsize of the {\em Herschel} SPIRE 250 $\mu$m observations (FWHM $\sim 17''$). Although many of the sources seem to be isolated field galaxies, we see that two sources (ALMA06 and ALMA07) have nearby infrared-bright companions that also have emission in CO(3-2). The detection of CO(3-2) at similar frequencies confirms that these galaxies are not simply isolated galaxies along the same line of sight, but are instead true close companions that will likely interact in the near future (the photometric redshifts of the companion galaxies are also well-matched with the central source's redshift). 

It can be a challenge to properly quantify the properties of these galaxies with close companions, as there are multiple IR-bright galaxies within the typical {\em Herschel} beam. This means that the $\lir$ that we measure in these systems is actually the luminosity from all of the components in the system, while the CO luminosities and stellar masses are resolved and can be measured independently. While ALMA06 and ALMA07 do not appear to be significant outliers in Figure \ref{fig:cutouts}, we will remove them from the remaining analysis when we calculate best-fit relationships describing galaxy properties. We should highlight the distinction between galaxies with close companions (ALMA06 \& ALMA07) and galaxies that are likely undergoing a merger (ALMA03 \& ALMA19).  The galaxies with neighbors are difficult to properly characterise because it is unclear how the far-infrared luminosity (and hence star formation rate) from {\em Herschel} should be split between the galaxies in the group, so we cannot measure the distance to the main sequence for any of the galaxies in the group.  On the other hand, ALMA03 and ALMA19 have a single stellar mass measurement and can be treated as a single system.  We include these galaxies in the following analysis, as they are star-forming galaxies and removing them would introduce a bias to our analysis. 

\section{Measuring Physical Properties}\label{sec:phys}

We move from an analysis of observed luminosities to a full analysis of the physical properties of galaxies. Due to the complications from combining disparate selection functions and observations from other studies, we focus the remaining analysis only on the sources in our survey, for which we understand the selection biases and observations. 

\subsection{Star Formation Rates}

Far-infrared luminosity is an excellent tracer of a galaxy's SFR as it directly probes the radiation from young, massive stars that is absorbed by interstellar dust and reradiated at infrared wavelengths. By combining our measured infrared luminosities with observations of the unobscured UV radiation in COSMOS from GALEX \citep{2007ApJS..172...99C}, we measure total star formation as $SFR_{\rm{Total}} = (8.6 \times 10^{-11}) \times (\lir + 2.3 \times \nu L_{\nu}(2300 \rm{\AA}))$, where $L_{\rm{IR}} \equiv L(8$--$1000 \mu$m) and all luminosities are measured in units of $L_{\odot}$ \citep{2013A&A...558A..67A}.  The infrared contribution generally dominates the emission, contributing at least 87\% of the total SFR in each source. 

\subsection{Gas Masses from CO}\label{sec:gas_mass}

The intensity of the ground rotational transition of carbon monoxide (CO(1-0)) probes $H_{2}$ column density via the relationship $N_{\mathrm{H}_{2}} = X_{\mathrm{CO}} \times I_{\mathrm{CO(1-0)}}$, or integrating over the emitting area and correcting for the mass contribution of heavier elements, $M_{mol} = \aco \times L_{\mathrm{CO}}'$, where $M_{mol}$ is the molecular gas mass and $\lco$ is the integrated line luminosity.  The conversion factor $\aco$ is the source of much uncertainty, but for the following analysis we adopt the metallicity dependent $\aco$ conversion factor from \citet{2015ApJ...800...20G}: $\alpha_{\mathrm{CO, Z}} = \alpha_{\mathrm{CO, MW}} \times \chi(\mathrm{Z})$. To maintain consistency, we adopt the parameters used in that work and set $\alpha_{\mathrm{CO, MW}} = 4.36~M_{\odot}/($K km s$^{-1}$ pc$^{2}$) and calculate $\chi(\mathrm{Z})$ as the geometric mean of the fitting functions from \citet{2013ARA&A..51..207B} and \citet{2012ApJ...746...69G}.  We convert metallicities from the \citet{2004ApJ...617..240K} scale to the \citet{2004MNRAS.348L..59P} scale using the calibrations in \citet{2008ApJ...681.1183K}. Sources without direct metallicity measurements have their metallicity estimated using the mass-metallicity relationship described in \citet{2015ApJ...800...20G}. We discuss the details and possible uncertainties of $\aco$ in Appendix \ref{sec:alpha_co}. 

\subsection{Star Formation Scaling Laws}

Stars form in giant molecular clouds that primarily contain molecular hydrogen, so we expect a strong correlation between galaxy gas masses and star formation rates. Indeed, surveys over a wide range of SFRs and surface densities show that the SFR per unit area follows the Kennicutt-Schmidt Law: 
\begin{equation}
\Sigma_{SFR} \propto \Sigma_{g}^{\beta}
\end{equation}
where $\Sigma_{SFR}$ is the star formation rate surface density, $\Sigma_{g}$ is the gas mass surface density, and the exponent ranges from $\beta = 1$--2, with a commonly adopted value of about 1.4 \citep{2012ARA&A..50..531K,2012ApJ...754....2S,2013AJ....146...19L}.

Recent studies have suggested that there may actually be two distinct and parallel correlations, separating normal star forming galaxies from starburst galaxies \citep{2007ApJ...671..303B,2010ApJ...714L.118D,2010MNRAS.407.2091G,2015ApJ...812L..23S}. But, this separation of normal and starburst galaxies may stem directly from the assumption of two very different $\aco$ factors for the two galaxy populations, and the two relationships may merge together when using a metallicity-dependent conversion factor \cite[e.g.][]{2011MNRAS.418..664N}
 
\begin{figure*}
\centering
\includegraphics[width=\textwidth, trim={60, 20, 120, 50}]{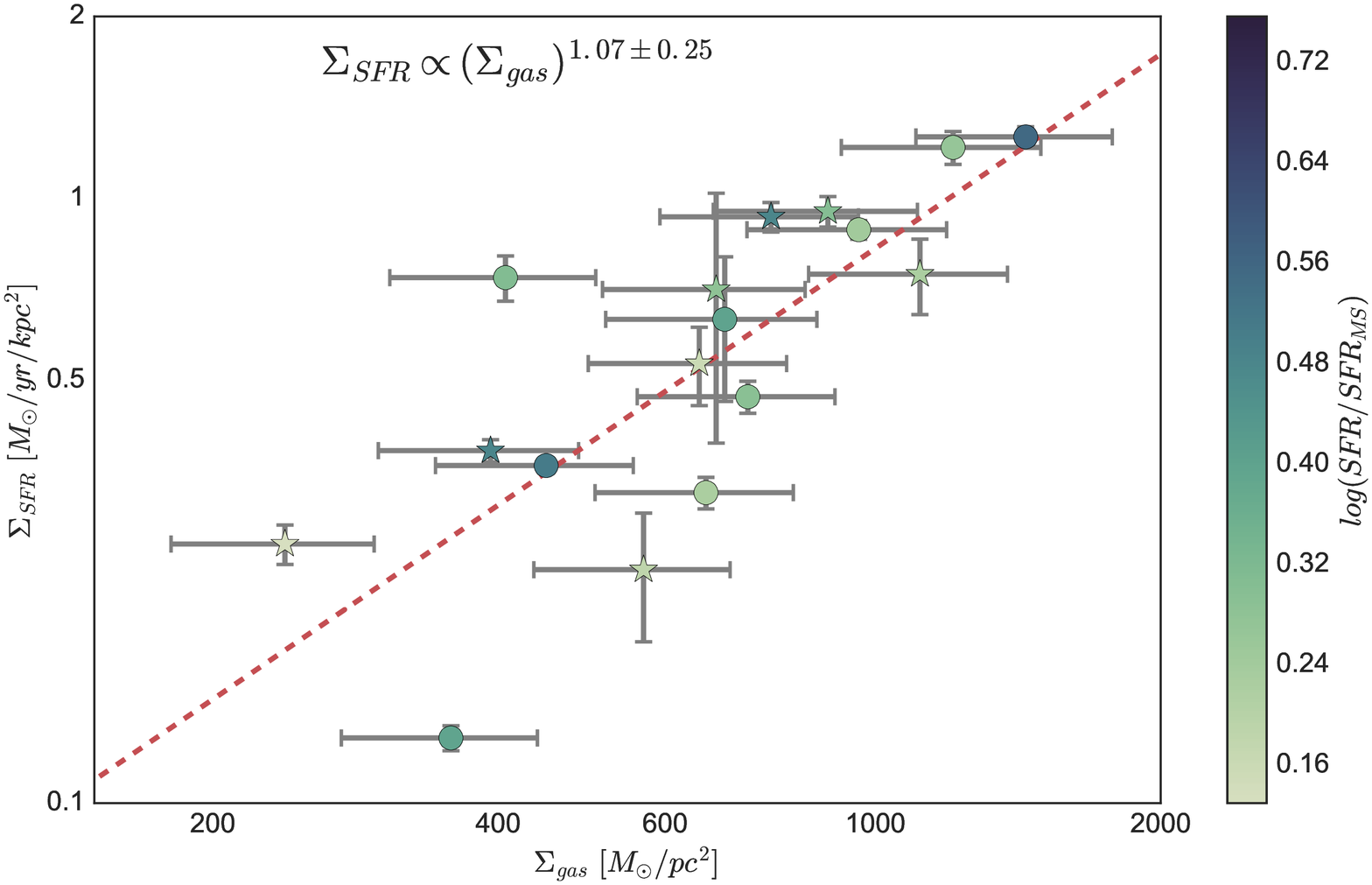}
\caption{Star formation rate surface density as a function of gas surface density.  Surface densities are calculated by dividing SFR and $M_{mol}$ by an area $A=\pi R_{half}^{2}$, where $R_{half}$ is the galaxy's half light radius as measured in the optical by {\em HST} ACS \citep{2009A&A...503..379T}.  The colours of the symobls represents the distance to the main sequence as indicated in the vertical colour bar. We fit a power-law of the form $\Sigma_{SFR} \propto \Sigma_{g}^{\beta}$ and find the best-fit relationship shown as the dashed red line.}
\label{fig:ks}
\end{figure*}

We calculate molecular gas masses using a metallicity dependent $\aco$ and plot the star formation rate density and molecular gas mass surface density of the galaxies in our sample in Figure \ref{fig:ks}. The colours of the circles represent each galaxy's distance to the main sequence. Surface densities have been calculated by dividing SFR and $M_{mol}$ by an area  $A=\pi R_{half}^{2}$, where $R_{half}$ is the galaxy's half light radius as measured in the optical by {\em HST} ACS \citep{2009A&A...503..379T}. The optical emission traces the total stellar mass distribution within a galaxy and does not necessarily correspond to the distribution of molecular gas or active star forming regions. However, our ALMA data are not spatially resolved, and the HST-based measurements provide the best size estimates for our sample.

Our sources show a very strong correlation between $\Sigma_{SFR}$ and $\Sigma_{gas}$, with a Spearman Rank correlation coefficient of $\rho = 0.83$ and $p = 8 \times 10^{-6}$ confirming clear correlation. We fit a power-law equation of the form $\Sigma_{SFR} \propto \Sigma_{g}^{\beta}$ to the data and find that the best-fit exponent is $\beta = 1.07 \pm 0.25$, which is consistent with the values of $\beta = 1$--2 seen in previous studies. We do not find evidence of two separate relationships in the $\Sigma_{SFR}-\Sigma_{gas}$ plane, although we are likely not sensitive to the extreme starburst galaxies that would populate the high SFE relationship. Within our sample, we do not see evidence that galaxies with a large distance to the main-sequence preferentially have higher than expected SFE.

\subsection{What drives $\Delta SFR_{MS}$?}

A galaxy's distance to the main sequence describes its SFR as compared to galaxies at similar stellar masses and redshifts. Galaxies with larger $\Delta SFR_{MS}$ have higher SFRs, and there are two likely explanations for this elevated SFR - an enhanced efficiency of converting a similar supply of molecular gas to stars, or simply a larger supply of the molecular gas needed to fuel star formation. With measurements of the molecular gas mass, we can now examine the extent that both of these factors drive the distance to the main sequence in the galaxies in our sample. 

\begin{figure*}
\centering
\includegraphics[width=\textwidth]{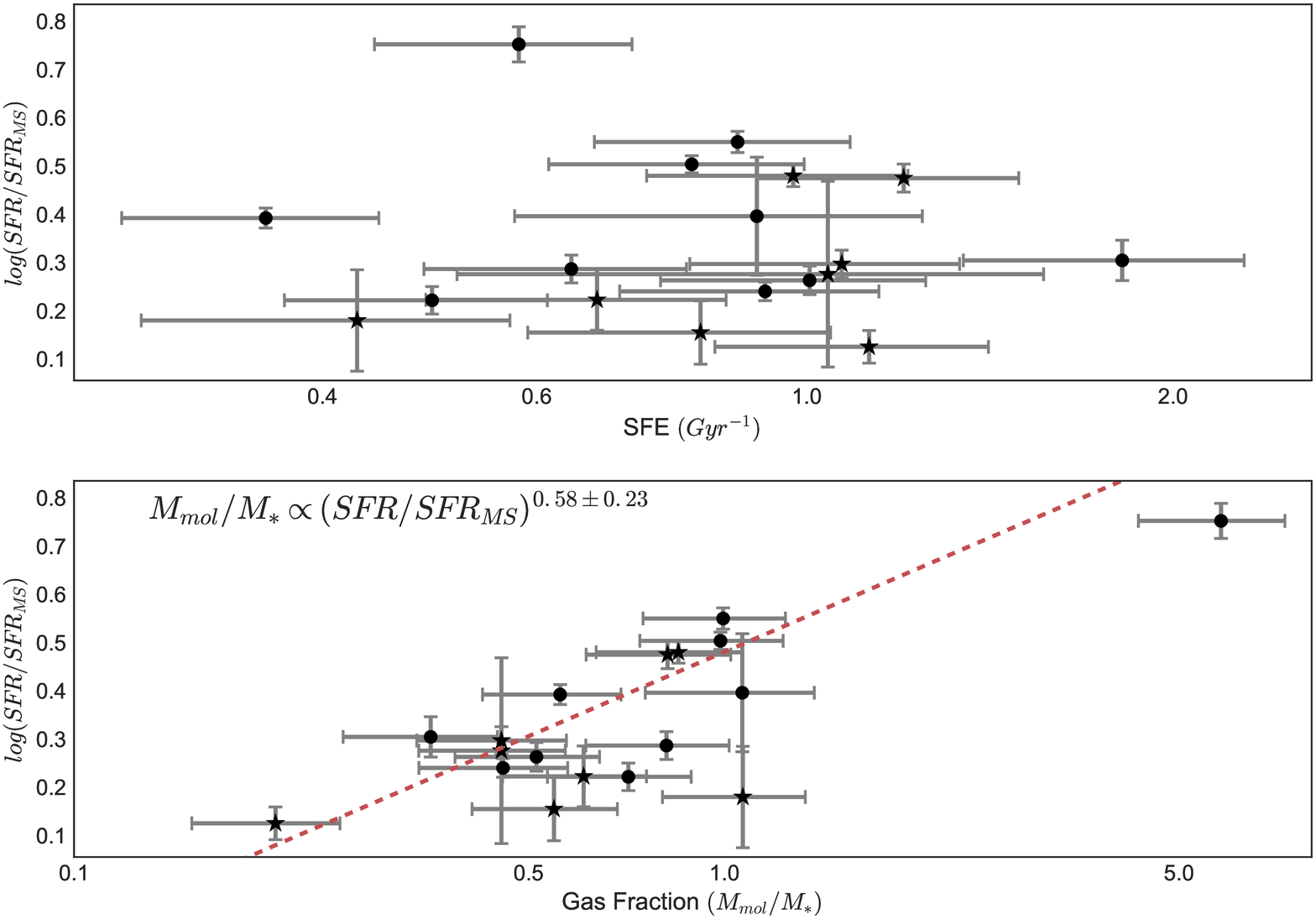}
\caption{A comparison of SFE (top panel) and gas fraction (bottom panel) with distance to the main sequence. Circles and stars again distinguish SFGs and AGN. We find no statistically significant correlation between SFE and distance to the main sequence, but we do find statistically significant correlation between gas fraction and distance to the main sequence (bottom). The best fit power-law relationship, $M_{mol} / M_{*} \propto (SFR/SFR_{MS})^{0.58 \pm 0.23}$, is plotted as the red dashed line.}
\label{fig:dist_ms}
\end{figure*}

Figure \ref{fig:dist_ms} displays how distance to the main sequence varies with star formation efficiency (top) and gas content (bottom). In the top panel of Figure \ref{fig:dist_ms}, we do not see any evidence that SFE and $\Delta SFR_{MS}$ are correlated in our sample. A Spearman Rank correlation test between the two parameters results in a correlation coefficient of $\rho = 0.133$ with a $p$-value of 0.2987, which cannot rule out the null hypothesis. In the bottom panel of Figure \ref{fig:dist_ms}, we compare the distance to the main sequence with molecular gas fraction ($f_{gas} = M_{mol}/M_{*}$). We see evidence of possible correlation between gas fraction and distance to the main sequence, and a Spearman Rank correlation test results in a correlation coefficient of $\rho = 0.546$ and a $p$-value of 0.0104, confirming that there is a statistically significant positive correlation. 

Taken together, the two plots in Figure \ref{fig:dist_ms} suggest that it is an increased gas supply, and not a change in star formation efficiency, that drives the $\Delta SFR_{MS}$ in our sample. 

\section{Discussion}\label{sec:discussion}

We find that an enhanced molecular gas content is likely driving the elevated star formation rates of the galaxies in our sample, while the star formation efficiency shows no systematic change with distance to the main sequence. Here we discuss the implications of these findings. 

In a study of $z \sim 0$ and $z \sim 2$ main sequence galaxies, \citet{2012ApJ...760....6M} found that the ``thickness'' of the main-sequence ($\Delta SFR_{MS} \lesssim 0.3$) is driven primarily by an evolving gas fraction, best parameterized as $M_{gas} / M_{*} \propto (sSFR/sSFR_{MS})^{\beta}$, with $\beta \sim 0.87$ for both local and high redshift samples.  We also find that an evolving gas fraction is primarily responsible for increased distance to the main-sequence, and we fit a relationship of $M_{gas} / M_{*} \propto (SFR/SFR_{MS})^{\beta}$ to our data and find a best-fit exponent of $\beta = 0.58 \pm 0.23$. This best-fit relationship is plotted in the bottom panel of Figure \ref{fig:dist_ms} as the red dashed line. Our measured slope is shallower than that found in \citet{2012ApJ...760....6M}, although this may be due to differences in sample selection, as our sample extends farther from the main-sequence out to $\Delta SFR_{MS} \lesssim 0.8$.  

In a study of over 600 galaxies with molecular gas masses derived from ALMA continuum observations, \citet{2017ApJ...837..150S} find that the molecular gas mass varies with distance to the main sequence as approximately $M_{ISM} \propto (sSFR/sSFR_{MS})^{0.38 \pm 0.06}$. This relationship is shallower than our derived fits, although the \citet{2017ApJ...837..150S} study is based on a sample of galaxies spanning a much larger dynamic range in both redshift and distance to the main sequence. Another difference is that \citet{2017ApJ...837..150S} also find that SFE increases with distance to the main sequence as $SFE \propto (sSFR/sSFR_{MS})^{0.60 \pm 0.10}$, while we do not see any correlation between SFE and $sSFR/sSFR_{MS}$. This could be explained if systematic SFE effects on distance to the main sequence only begin to play a significant role for galaxies significantly offset from the main sequence, as suggested in \citet{2012ApJ...760....6M,2014ApJ...793...19S}.

We note that while our sample extends out to $\Delta SFR_{MS} \lesssim 0.8$, the majority of our sources are actually at $\Delta SFR_{MS} \lesssim 0.6$, with only one source with a higher distance to the main sequence, so we are limited in sample size and are unable to draw strong conclusions for galaxies in the range $0.6 \lesssim \Delta SFR_{MS} \lesssim 0.8$.  In particular, the galaxy with the highest distance to the main sequence, ALMA19 ($\Delta SFR_{MS} = 0.78$), is the galaxy identified as a possible post-merger system from its morphology (see Figure \ref{fig:cutouts}). Removing this galaxy from our samples does not change the correlations found (or not found) in Figure \ref{fig:dist_ms}.

While we do not see evidence that SFE is correlated with distance to the main sequence, our sample is not sensitive to the most extreme outliers to the main sequence with $\Delta SFR >1$  ($SFR/SFR_{MS} > 10$) that may still be undergoing a different mode of star formation. As seen in Section \ref{sec:lir_lco}, the $\lir / \lco$ ratios of these extreme starburst galaxies suggest they may indeed have increased star formation efficiencies, although there are selection effects keep in mind. 

Although we are not sensitive to extreme outliers to the main sequence, we find that gas fraction is the dominant factor for the enhanced SFR in galaxies with $\Delta SFR_{MS} \lesssim 0.6$, and that the relationship between gas fraction and $\Delta SFR_{MS}$ is consistent with what has been found in galaxies much closer to the main sequence. Taken together with the morphology of these galaxies discussed in Section \ref{sec:morph} (predominantly disk-like), it is likely that the galaxies in this ``transition'' region are similar to main sequence galaxies with high gas fractions.  We do not see any systematic evolution to a different mode of star formation with drastically increased efficiency in our sample, and the SFE and distance to the main sequence are statistically uncorrelated in our sample.

\section{Summary}

We study the CO $J=3\rightarrow2$ luminosity in a sample of 20 LIRGs at redshifts $z = 0.25$--0.65. By taking advantage of the excellent multi-wavelength coverage of COSMOS, we accurately measure key properties that allow us to probe the nature of star formation in these galaxies. 

Due to the many assumptions required to determine molecular gas masses from CO $J=3\rightarrow2$ luminosity, we first focus on the observed quantities $\lco$ and $\lir$. Our LIRGs fall in the locus of most other galaxies with similar luminosities, but do not show any sign of scattering to ``starburst'' tracks found in the literature. The $\lir / \lco$ ratio does not appear to depend on distance to main sequence or AGN presence, although a comparison to other studies at similar redshift ranges suggests that there may be an increase in $\lir / \lco$ ratio at very high $\Delta SFR_{MS}$. 

We then use a metallicity-dependent $\aco$ conversion factor to calculate molecular gas masses and examine the physical properties of our galaxies. We find that:
 \begin{itemize}
 \item The majority of the sample has disk-like morphologies, with only $\sim 10$\% showing signs of interaction. The interacting galaxies generally have high distance to the main sequence, but do not have preferentially high SFE. 
 \item Distance to the main sequence and SFE are not statistically correlated, suggesting that an increased star formation efficiency (or different mode of star formation) does not play a major role in the elevated SFR in galaxies at $\Delta SFR_{MS} \lesssim 0.6$. 
 \item Gas fraction and distance to the main sequence are statistically correlated, and we find a best-fit relationship of $M_{gas} / M_{*} \propto (sSFR/sSFR_{MS})^{0.58 \pm 0.23}$. 
 \end{itemize}

Taken together, all of these results suggest that galaxies with $\Delta SFR_{MS}  \lesssim 0.6$ are likely still normal, main sequence galaxies that have elevated star formation rates due to their larger molecular gas content.  At these moderate distances to the main sequence, starburst galaxies likely do not make up a significant fraction of the population, and it is instead the most extreme outliers to the main sequence that are likely undergoing a separate mode of star formation.

\section{Acknowledgements}

We would like to thank the staff at the Nordic Regional ALMA Centre for all of their help in reducing and re-reducing the problematic Cycle 1 ALMA data. 
This publication has received funding from the European Union's Horizon 2020 research and innovation programme under grant agreement No 730562 [RadioNet]. 

COSMOS is based on observations with the NASA/ESA {\em Hubble Space
Telescope}, obtained at the Space Telescope Science Institute, which
is operated by AURA Inc, under NASA contract NAS 5-26555; also based
on data collected at: the Subaru Telescope, which is operated by the
National Astronomical Observatory of Japan; the XMM-Newton, an ESA
science mission with instruments and contributions directly funded by
ESA Member States and NASA; the European Southern Observatory, Chile;
Kitt Peak National Observatory, Cerro Tololo Inter-American
Observatory, and the National Optical Astronomy Observatory, which are
operated by the Association of Universities for Research in Astronomy,
Inc. (AURA) under cooperative agreement with the National Science
Foundation; the National Radio Astronomy Observatory which is a
facility of the National Science Foundation operated under cooperative
agreement by Associated Universities, Inc; and the
Canada-France-Hawaii Telescope operated by the National Research
Council of Canada, the Centre National de la Recherche Scientifique de
France and the University of Hawaii.

N.L, S.T, G.E.M, and I.C acknowledge  support from the ERC Consolidator Grant funding scheme (project ConTExt, grant No. 648179). C.\ M.\ Casey thanks the UT Austin College of Natural Science for support. G.E.M acknowledges support from the Carlsberg Foundation, the ERC Consolidator Grant funding scheme (project ConTExt, grant number No. 648179), and a research grant (13160) from Villum Fonden. AK acknowledges support by the Collaborative Research Center 956, sub-project A1, funded by the Deutsche Forschungsgemeinschaft (DFG).

KS and KS acknowledge support from the North American ALMA Science Center at the National Radio Astronomy Observatory which is a facility of the National Science Foundation operated under cooperative agreement by Associated Universities, Inc. This paper makes use of the following ALMA data: ADS/JAO.ALMA\#2012.1.00076. ALMA is a partnership of ESO (representing its member states), NSF (USA) and NINS (Japan), together with NRC (Canada), NSC and ASIAA (Taiwan), and KASI (Republic of Korea), in cooperation with the Republic of Chile. The Joint ALMA Observatory is operated by ESO, AUI/NRAO and NAOJ.




\bibliographystyle{mnras}
\bibliography{apj-jour,astrobib}







\appendix

\section{Measuring Gas Masses}

\subsection{CO Excitation} \label{sec:r31}

CO line emission has been the most widely used tracer of the molecular gas mass, but only CO $J=1\rightarrow0$ luminosities have been well correlated with the virial masses of self-gravitating GMCs in the Milky Way \citep{1987ApJS...63..821S,1987ApJ...319..730S}. Observed luminosities of higher $J$ transitions of CO must be converted to CO $J=1\rightarrow0$ through assumptions on the CO excitation. The CO excitation is determined by the temperature and density of the gas, which can vary quite dramatically between individual galaxies, but the average excitation ladder for low $J$ transitions is fairly consistent  \citep[e.g.][]{2013ARA&A..51..105C,2016ApJ...829...93K}. 

In our study, we concentrate on the relative strengths of the CO $J=1\rightarrow0$ and CO $J=3\rightarrow2$ rotational transitions. Often, luminosities of higher excitation lines are converted to $L_{\mathrm{CO(1-0)}}'$ using conversion factors $R_{J1} \equiv L_{\mathrm{CO}(J-(J-1))}' / L_{\mathrm{CO(1-0)}}'$.  \citet{2013ARA&A..51..105C} find that $R_{31}$ can vary from 0.27 (Milky Way) to 0.97 (quasars).  However, studies of typical star forming galaxies and starburst galaxies find a fairly consistent average value of $<R_{31}> \sim 0.5$--0.65 \citep{2013ARA&A..51..105C,2014ApJ...794..142G,2015ApJ...800...20G}.

\begin{figure}
\centering
\includegraphics[width=\columnwidth]{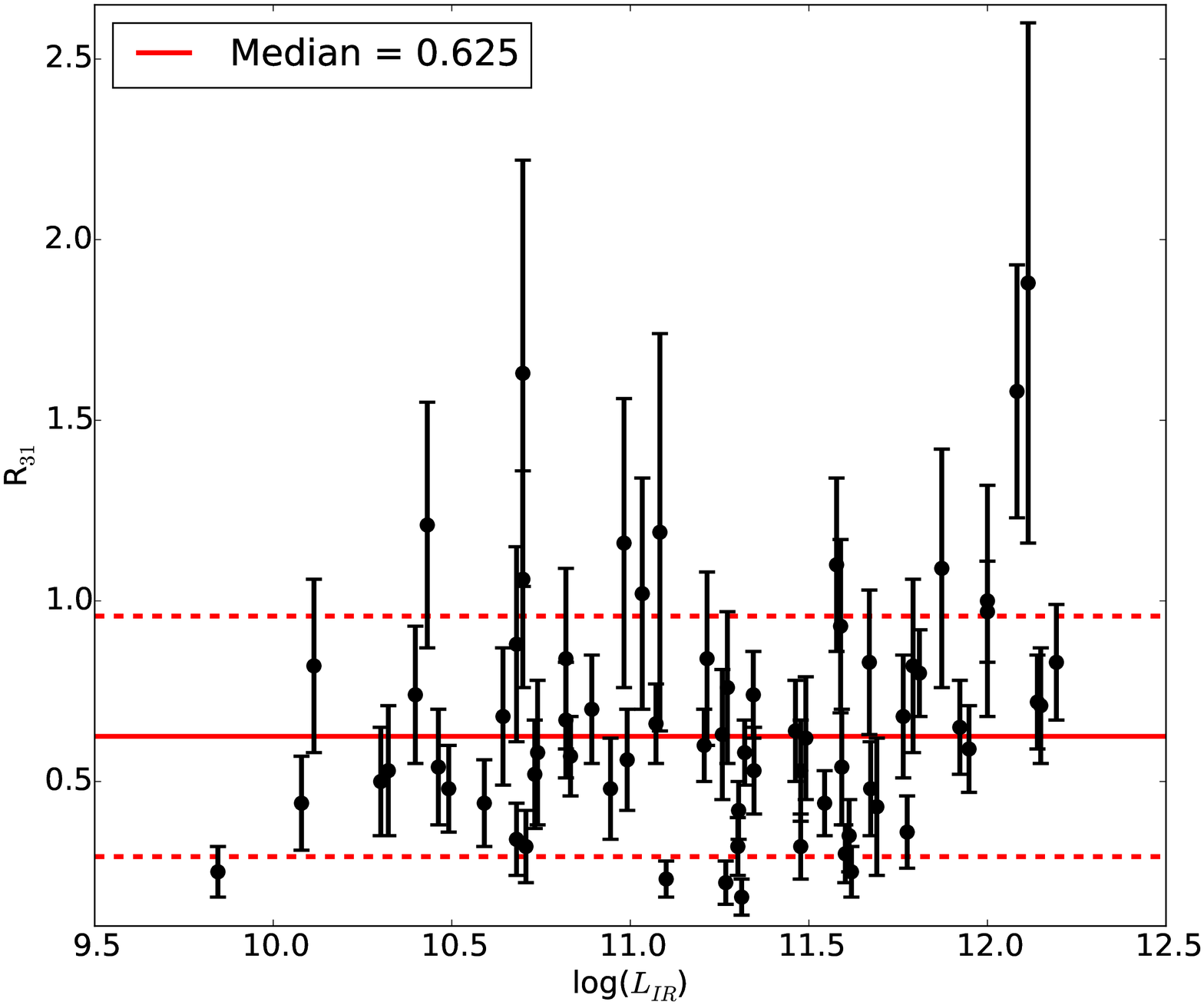}
\caption{$R_{31}$ values of local (U)LIRGs in \citet{2012MNRAS.426.2601P}. The red solid line represents the median value of $R_{31}$ across the whole sample, and the dashed red lines represent the standard deviation. }
\label{fig:r31}
\end{figure}

In this paper, we use the median value of $R_{31} = 0.625 \pm 0.335$ found in the study of \citet{2012MNRAS.426.2601P}, who studied the CO excitation in the HerCULES survey of 70 infrared luminous galaxies in the local universe (see Figure \ref{fig:r31}). The infrared luminosities sampled in the HerCULES sample span log($\lir) \sim 10$--12, with a median log$(\lir) = 11.2$, while the 20 galaxies in our sample span log($\lir) = 11.05$--11.66 with a median log$(\lir) = 11.39$. Thus, the HerCULES data provides a relatively well-matched sample in luminosity, but covers a much wider range of galaxies. We adopt the standard deviation of $R_{31}$ in the HerCULES sample as an uncertainty that is propagated through all of our calculations involving $ L_{\mathrm{CO(1-0)}}'$. Given the much wider luminosity range of galaxies probed in HerCULES, this is likely an overestimate of the variation in $R_{31}$, but provides a conservative measure of the uncertainty.

The use of an average CO excitation may be less accurate in AGN host galaxies, as AGN have been demonstrated to drive mechanical or radiative feedback that can heat molecular gas reservoirs and lead to higher molecular line excitation \citep[e.g.][]{2008A&A...491..483P,2010A&A...518L..42V}, and quasar hosts at high redshift show the highest molecular gas excitation \citep{1997ApJ...484..695B,2009ApJ...703.1338R}. However, we decide to apply the mean $R_{31}$ value from \citet{2012MNRAS.426.2601P} to the AGN in our sample (as well as the SFGs) because $(i)$ our AGN-selection from strong-emission lines is incomplete because only half of our sources have Hectospec spectra, and $(ii)$ the \citet{2012MNRAS.426.2601P} parent sample is infrared selected and includes many AGN hosts, which suggests that the average we derive encompasses the extra excitation from AGN.  

\subsection{The CO-to-H$_{2}$ conversion}\label{sec:alpha_co}

Bolatto et al.\ (2013) provide a summary of the numerous studies of the conversion factor between $L_{\mathrm{CO(1-0)}}'$ and $M_{mol}$ and conclude that a fairly consistent value of $\aco \sim 4~M_{\odot}/($K km s$^{-1}$ pc$^{2}$) is appropriate for giant molecular clouds in the Milky Way and nearby spiral galaxies. \citet{2016ApJ...820...83S} argue for a slightly higher value of $\aco = 6.5~ M_{\odot}/($K km s$^{-1}$ pc$^{2}$) based on measurements of over 500 resolved GMCs in the inner galactic plane \citep{1987ApJS...63..821S,1987ApJ...319..730S}.

However, in nearby nuclear starbursts, it was found that the use of the MW conversion factor implied molecular gas masses that exceeded the dynamical mass \citep{1999AJ....117.2632B}.  \citet{1998ApJ...507..615D} found a characteristic value of $\aco \sim 0.8~M_{\odot}/($K km s$^{-1}$ pc$^{2}$) is more appropriate for these sources. Common practice is to apply a bimodal $\aco$, assuming a MW value for normal star forming galaxies and this significantly reduced value for starburst galaxies.  Theoretically, the value of the CO-to-H$_{2}$ conversion factor  is expected to vary as $\sqrt{n_{\mathrm{H_{2}}}} / T_{k}$, and so may be expected to be different in the dense, hot environments of a nuclear starburst.  However, \citet{2016ApJ...820...83S} suggest that while this lower conversion factor may be appropriate for studies of isolated nuclear regions, studies of a galaxy's overall gas content should still use a Milky-Way like conversion factor because the bulk of the gas will not be in these extreme environments, even in starburst galaxies \citep[also see discussion in][]{2015ApJ...800...70S,2016arXiv160509381S}. 

While many studies suggest that a single (or bimodal) value of $\aco$ is appropriate for the bulk of galaxies, some argue that a continuously varying $\aco$ may be a more accurate description of the CO-to-H$_{2}$ conversion. In particular, metallicity is expected to have a significant effect on $\aco$, especially at low metallicities \citep[e.g.][]{2011ApJ...737...12L,2012ApJ...746...69G,2013ARA&A..51..207B}. In this paper, we apply the metallicity-dependent $\aco$ described in \citet{2015ApJ...800...20G}: $\alpha_{\mathrm{CO, Z}} = \alpha_{\mathrm{CO, MW}} \times \chi(\mathrm{Z})$ (see Section \ref{sec:gas_mass}). 

\subsection{Gas Masses from Dust Continuum}\label{sec:gas_dust}

Gas masses can also be measured from emission in the far-infrared continuum if the dust emissivity and dust-to-gas ratio are known \citep[e.g.][]{2011ApJ...740L..15M,2011A&A...536A..19P}. Gas masses measured from dust continuum usually probe the {\em total} gas mass (atomic \& molecular), while CO probes only the molecular gas mass. Studies suggest that the molecular gas mass is much more strongly correlated with star formation \citep[see review in][]{2012ARA&A..50..531K}, so often we are more interested in the molecular gas mass. This means that dust-based measurements of the molecular gas mass are subject to an additional uncertainty in the HI gas mass. 

\citet{2014ApJ...783...84S,2016ApJ...820...83S} empirically calibrate restframe 850 $\mu$m luminosity with {\em molecular} gas mass and find a strikingly constant value of $L_{\nu_{850 \mu{\mathrm m}}} / M_{\mathrm{mol}} = 6.7 \times 10^{19} \mathrm{ergs\ sec}^{-1} \mathrm{Hz}^{-1} M_{\odot}^{-1}$ across a wide range of galaxies. This calibration provides an independent measure of molecular gas mass from the far-infrared continuum photometry. Thus, we can also calculate molecular gas masses from observed {\em Herschel} SPIRE 350 $\mu$m flux densities (where available) using Equation A14 in \citet{2016ApJ...820...83S}: 
\begin{multline*}
M_{\mathrm{mol}} = 1.78 S_{\nu_{\mathrm{obs}}} [\mathrm{mJy}] (1+z)^{-4.8} \\
\times \big(\frac{\nu_{850 \mu\mathrm{m}}}{\nu_{\mathrm{obs}}}\big)^{3.8} (d_{L} [\mathrm{Gpc}])^{2} \frac{\Gamma_{0}}{\Gamma_{RJ}}\ 10^{10} M_{\odot}
\end{multline*}
 where the $\Gamma$ factors are corrections for departure of the Planck function from the Rayleigh-Jeans. Although this estimate of molecular gas mass from dust continuum avoids the uncertain contribution from HI gas masses, this is still subject to uncertainties on the dust emissivity and dust-to-gas ratio, which can vary with galaxy properties such as metallicity \citep[e.g.][]{2007ApJ...663..866D,2014A&A...563A..31R}.

\subsection{Comparison of Gas Masses}\label{sec:gas_mass_comp}

We calculate molecular gas masses for our galaxies using all of the above described methods for measuring gas mass, and find that a single-value MW $\aco$, a metallicity dependent $\aco$, and dust-continuum estimates all result in a consistent measurement of the gas mass, with typical differences of $\lesssim 0.05$ dex. This provides confidence that the results in our paper are robust to the particular method used to estimate gas masses. 

The one method that could lead to significant changes would be assumption of a bimodal $\aco$, where one assumes a drastically lower conversion factor for a subset of the sample. Presumably the lower $\alpha_{\mathrm{CO,SB}}$ factor would be applied to galaxies above a certain $\Delta SFR_{MS}$ threshold, and the choice of this threshold would also affect the results.  

The assumption of a bimodal $\aco$ would lead to much lower gas masses in the galaxies above the $\Delta SFR_{MS}$ threshold and could reverse the trends seen in Figure \ref{fig:dist_ms}, such that SFE would show some correlation with distance to the main sequence and gas fraction would no longer have such a strong correlation.  However, these new relationships would be step-functions created solely from the assumption of two very different conversion factors. None of the trends seen in the analysis of the directly observable quantities $\lir$ and $\lco$ suggest the need for two different conversion factors in our sample, and the introduction of a bimodal conversion factor would lead to relationships that differ significantly from those seen in the directly observed values. These all suggest that a bimodal $\aco$ is inappropriate, at least for the galaxies in our sample with $\Delta SFR_{MS} \lesssim 0.6$--0.8. 

\bsp	
\label{lastpage}
\end{document}